\documentclass{iopconfser}

\usepackage{amsthm}
\usepackage{amsmath,amssymb}
 \usepackage{graphicx}
 \usepackage{booktabs}
\usepackage{subcaption}  
\usepackage{caption}
\usepackage{float} % For figure/table positioning
\usepackage{xspace}
\usepackage{color}
\usepackage{epsfig}
\graphicspath{{.}{./figures/}}
\usepackage{wrapfig}
\usepackage{macros/tikzfig}
\usepackage{keycommand}

\usepackage{lineno}
 \usepackage{hyperref}
 \usepackage{wrapfig}

\newkeycommand{\boxmap}[style=small box][1]{\,\tikz{\node[style=\commandkey{style}] (x) {$#1$};\draw(0,-1.25)--(x)--(0,1.25);}\,}
\newkeycommand{\dboxmap}[style=dbox][1]{\,\tikz{\node[style=\commandkey{style}] (x) {$#1$};\draw[boldedge] (0,-1.25)--(x)--(0,1.25);}\,}

\newkeycommand{\wirelabel}[style=white label]{\,\tikz{\node[style=\commandkey{style}]{};}\,\xspace}

\newkeycommand{\pointmap}[style=point][1]{\,\tikz{\node[style=\commandkey{style}] (x) at (0,-0.3) {$#1$};\node [style=none] (3) at (0, 0.9) {};\node [style=none] (3) at (0, -0.9) {};\draw (x)--(0,0.7);}\,}

\newkeycommand{\point}[style=point,estyle=][1]{\,\tikz{\node[style=\commandkey{style}] (x) at (0,-0.05) {$#1$};\draw [\commandkey{estyle}] (x)--(0,1.05);}\,}

\newkeycommand{\copointmap}[style=copoint][1]{\,\tikz{\node[style=\commandkey{style}] (x) at (0,0.3) {$#1$};\draw (x)--(0,-0.7);}\,}

\newkeycommand{\dpoint}[style=point][1]{\,\tikz{\node[style=\commandkey{style},doubled] (x) at (0,-0.3) {$#1$};\draw[boldedge] (x)--(0,0.7);}\,}

\newkeycommand{\kpoint}[style=kpoint,estyle=][1]{\,\tikz{\node[style=\commandkey{style}] (x) at (0,-0.05) {$#1$};\draw [\commandkey{estyle}] (x)--(0,1.05);}\,}

\newkeycommand{\kpointgray}[style=gray kpoint,estyle=][1]{\,\tikz{\node[style=\commandkey{style}] (x) at (0,-0.05) {$#1$};\draw [\commandkey{estyle}] (x)--(0,1.05);}\,}

\newkeycommand{\typedkpoint}[style=kpoint,edgestyle=][2]{%
\,\begin{tikzpicture}
  \begin{pgfonlayer}{nodelayer}
    \node [style=none] (0) at (0, 1) {};
    \node [style=\commandkey{style}] (1) at (0, -0.75) {$#2$};
    \node [style=right label] (2) at (0.25, 0.5) {$#1$};
  \end{pgfonlayer}
  \begin{pgfonlayer}{edgelayer}
    \draw [\commandkey{edgestyle}] (1) to (0.center);
  \end{pgfonlayer}
\end{tikzpicture}\,}

\newkeycommand{\typedkpointdag}[style=kpoint adjoint,edgestyle=][2]{%
\,\begin{tikzpicture}
  \begin{pgfonlayer}{nodelayer}
    \node [style=none] (0) at (0, -1) {};
    \node [style=\commandkey{style}] (1) at (0, 0.75) {$#2$};
    \node [style=right label] (2) at (0.25, -0.5) {$#1$};
  \end{pgfonlayer}
  \begin{pgfonlayer}{edgelayer}
    \draw [\commandkey{edgestyle}] (0.center) to (1);
  \end{pgfonlayer}
\end{tikzpicture}\,}

\newkeycommand{\bistate}[style=kpoint][1]{\,\begin{tikzpicture}
  \begin{pgfonlayer}{nodelayer}
    \node [style=\commandkey{style}, minimum width=1 cm, inner sep=2pt] (0) at (0, -0.25) {$#1$};
    \node [style=none] (1) at (-0.75, 0) {};
    \node [style=none] (2) at (0.75, 0) {};
    \node [style=none] (3) at (-0.75, 0.88) {};
    \node [style=none] (4) at (0.75, 0.88) {};
  \end{pgfonlayer}
  \begin{pgfonlayer}{edgelayer}
    \draw (1.center) to (3.center);
    \draw (2.center) to (4.center);
  \end{pgfonlayer}
\end{tikzpicture}\,}

\newkeycommand{\bistatebig}[style=kpoint][1]{\,\begin{tikzpicture}
  \begin{pgfonlayer}{nodelayer}
    \node [style=\commandkey{style}, minimum width=, minimum width=1.26 cm, inner sep=2pt] (0) at (0, -0.25) {$#1$};
    \node [style=none] (1) at (-1, 0) {};
    \node [style=none] (2) at (1, 0) {};
    \node [style=none] (3) at (-1, 0.88) {};
    \node [style=none] (4) at (1, 0.88) {};
  \end{pgfonlayer}
  \begin{pgfonlayer}{edgelayer}
    \draw (1.center) to (3.center);
    \draw (2.center) to (4.center);
  \end{pgfonlayer}
\end{tikzpicture}\,}

\newkeycommand{\dbistate}[style=dkpoint][1]{\,\begin{tikzpicture}
  \begin{pgfonlayer}{nodelayer}
    \node [style=\commandkey{style}, minimum width=1 cm, inner sep=2pt] (0) at (0, -0.25) {$#1$};
    \node [style=none] (1) at (-0.75, 0) {};
    \node [style=none] (2) at (0.75, 0) {};
    \node [style=none] (3) at (-0.75, 1.0) {};
    \node [style=none] (4) at (0.75, 1.0) {};
  \end{pgfonlayer}
  \begin{pgfonlayer}{edgelayer}
    \draw [boldedge] (1.center) to (3.center);
    \draw [boldedge] (2.center) to (4.center);
  \end{pgfonlayer}
\end{tikzpicture}\,}

\newkeycommand{\bistatebraket}[style1=kpoint,style2=kpointadj][2]{\,\begin{tikzpicture}
  \begin{pgfonlayer}{nodelayer}
    \node [style=\commandkey{style1}, minimum width=1 cm, inner sep=2pt] (0) at (0, -0.75) {$#1$};
    \node [style=none] (1) at (-0.75, -0.5) {};
    \node [style=none] (2) at (0.75, -0.5) {};
    \node [style=none] (3) at (-0.75, 0.5) {};
    \node [style=none] (4) at (0.75, 0.5) {};
    \node [style=\commandkey{style2}, minimum width=1 cm, inner sep=2pt] (5) at (0, 0.75) {$#2$};
  \end{pgfonlayer}
  \begin{pgfonlayer}{edgelayer}
    \draw (1.center) to (3.center);
    \draw (2.center) to (4.center);
  \end{pgfonlayer}
\end{tikzpicture}\,}

\newkeycommand{\weight}[style=dkpoint][1]{\,\begin{tikzpicture}
  \begin{pgfonlayer}{nodelayer}
    \node [style=\commandkey{style}] (0) at (0, -0.5) {$#1$};
    \node [style=upground] (1) at (0, 0.75) {};
  \end{pgfonlayer}
  \begin{pgfonlayer}{edgelayer}
    \draw [style=boldedge] (0) to (1);
  \end{pgfonlayer}
\end{tikzpicture}\,}

\newkeycommand{\dkpoint}[style=dkpoint][1]{\,\tikz{\node[style=\commandkey{style}] (x) at (0,-0.1) {$#1$};\draw[boldedge] (x)--(0,1);}\,}
\newkeycommand{\dkpointadj}[style=dkpointadj][1]{\,\tikz{\node[style=\commandkey{style}] (x) at (0,0.1) {$#1$};\draw[boldedge] (x)--(0,-1);}\,}
\newkeycommand{\dkpointtrans}[style=dkpointtrans][1]{\,\tikz{\node[style=\commandkey{style}] (x) at (0,0.1) {$#1$};\draw[boldedge] (x)--(0,-1);}\,}
\newkeycommand{\dkpointconj}[style=dkpointconj][1]{\,\tikz{\node[style=\commandkey{style}] (x) at (0,-0.1) {$#1$};\draw[boldedge] (x)--(0,1);}\,}

\newkeycommand{\blackkpoint}[style=black kpoint][1]{\,\tikz{\node[style=\commandkey{style}] (x) at (0,-0.1) {$#1$};\draw (x)--(0,1);}\,}
\newkeycommand{\blackkpointadj}[style=black kpointadj][1]{\,\tikz{\node[style=\commandkey{style}] (x) at (0,0.1) {$#1$};\draw (x)--(0,-1);}\,}
\newkeycommand{\blackdkpoint}[style=black dkpoint][1]{\,\tikz{\node[style=\commandkey{style}] (x) at (0,-0.1) {$#1$};\draw[boldedge] (x)--(0,1);}\,}
\newkeycommand{\blackdkpointadj}[style=black dkpointadj][1]{\,\tikz{\node[style=\commandkey{style}] (x) at (0,0.1) {$#1$};\draw[boldedge] (x)--(0,-1);}\,}

\newcommand{\trace}{\,\begin{tikzpicture}
  \begin{pgfonlayer}{nodelayer}
    \node [style=none] (0) at (0, -0.60) {};
    \node [style=upground] (1) at (0, 0.40) {};
  \end{pgfonlayer}
  \begin{pgfonlayer}{edgelayer}
    \draw [style=none] (0.center) to (1);
  \end{pgfonlayer}
\end{tikzpicture}\,}

\newcommand\discard\trace

\newkeycommand{\pointketbra}[style1=copoint,style2=point,estyle=][2]{\,%
\begin{tikzpicture}
  \begin{pgfonlayer}{nodelayer}
    \node [style=none] (0) at (0, -1.75) {};
    \node [style=\commandkey{style1}] (1) at (0, -0.75) {$#1$};
    \node [style=\commandkey{style2}] (2) at (0, 0.75) {$#2$};
    \node [style=none] (3) at (0, 1.75) {};
  \end{pgfonlayer}
  \begin{pgfonlayer}{edgelayer}
    \draw [\commandkey{estyle}] (0.center) to (1);
    \draw [\commandkey{estyle}] (2) to (3.center);
  \end{pgfonlayer}
\end{tikzpicture}\,}

\newkeycommand{\twocopointketbra}[style1=copoint,style2=copoint,style3=point][3]{\,%
\begin{tikzpicture}
  \begin{pgfonlayer}{nodelayer}
    \node [style=none] (0) at (-0.75, -1.75) {};
    \node [style=none] (1) at (0.75, -1.75) {};
    \node [style=\commandkey{style1}] (2) at (-0.75, -0.75) {$#1$};
    \node [style=\commandkey{style2}] (3) at (0.75, -0.75) {$#2$};
    \node [style=\commandkey{style3}] (4) at (0, 0.75) {$#3$};
    \node [style=none] (5) at (0, 1.75) {};
  \end{pgfonlayer}
  \begin{pgfonlayer}{edgelayer}
    \draw (0.center) to (2);
    \draw (1.center) to (3);
    \draw (4) to (5.center);
  \end{pgfonlayer}
\end{tikzpicture}\,}

\newkeycommand{\twopointketbra}[style1=copoint,style2=point,style3=point][3]{\,%
\begin{tikzpicture}
  \begin{pgfonlayer}{nodelayer}
    \node [style=none] (0) at (0, -1.75) {};
    \node [style=none] (1) at (-0.75, 1.75) {};
    \node [style=\commandkey{style1}] (2) at (0, -0.75) {$#1$};
    \node [style=\commandkey{style2}] (3) at (-0.75, 0.75) {$#2$};
    \node [style=\commandkey{style3}] (4) at (0.75, 0.75) {$#3$};
    \node [style=none] (5) at (0.75, 1.75) {};
  \end{pgfonlayer}
  \begin{pgfonlayer}{edgelayer}
    \draw (0.center) to (2);
    \draw (1.center) to (3);
    \draw (4) to (5.center);
  \end{pgfonlayer}
\end{tikzpicture}\,}

\newkeycommand{\pointbraket}[style1=point,style2=copoint,estyle=][2]{\,%
\begin{tikzpicture}
  \begin{pgfonlayer}{nodelayer}
    \node [style=\commandkey{style1}] (0) at (0, -0.625) {$#1$};
    \node [style=\commandkey{style2}] (1) at (0, 0.625) {$#2$};
  \end{pgfonlayer}
  \begin{pgfonlayer}{edgelayer}
    \draw [\commandkey{estyle}] (0) to (1);
  \end{pgfonlayer}
\end{tikzpicture}\,}

\newkeycommand{\kpointketbra}[style1=kpointdag,style2=kpoint,estyle=][2]{\,%
\begin{tikzpicture}
  \begin{pgfonlayer}{nodelayer}
    \node [style=none] (0) at (0, -1.9) {};
    \node [style=\commandkey{style1}] (1) at (0, -0.95) {$#1$};
    \node [style=\commandkey{style2}] (2) at (0, 0.95) {$#2$};
    \node [style=none] (3) at (0, 1.9) {};
  \end{pgfonlayer}
  \begin{pgfonlayer}{edgelayer}
    \draw [\commandkey{estyle}] (0.center) to (1);
    \draw [\commandkey{estyle}] (2) to (3.center);
  \end{pgfonlayer}
\end{tikzpicture}\,}

\newkeycommand{\kpointmap}[style1=kpoint,style2=map][2]{\,%
\begin{tikzpicture}
  \begin{pgfonlayer}{nodelayer}
    \node [style=\commandkey{style1}] (0) at (0, -1.1) {$#1$};
    \node [style=\commandkey{style2}] (1) at (0, 0.5) {$#2$};
    \node [style=none] (2) at (0, 1.75) {};
  \end{pgfonlayer}
  \begin{pgfonlayer}{edgelayer}
    \draw (0) to (1);
    \draw (1) to (2.center);
  \end{pgfonlayer}
\end{tikzpicture}%
\,}

\newkeycommand{\kpointmaptrans}[style1=kpointconj,style2=mapadj][2]{\,%
\begin{tikzpicture}
  \begin{pgfonlayer}{nodelayer}
    \node [style=\commandkey{style1}] (0) at (0, -1.1) {$#1$};
    \node [style=\commandkey{style2}] (1) at (0, 0.5) {$#2$};
    \node [style=none] (2) at (0, 1.75) {};
  \end{pgfonlayer}
  \begin{pgfonlayer}{edgelayer}
    \draw (0) to (1);
    \draw (1) to (2.center);
  \end{pgfonlayer}
\end{tikzpicture}%
\,}

\newkeycommand{\kpointmapadj}[style1=kpointadj,style2=map][2]{\,%
\begin{tikzpicture}
  \begin{pgfonlayer}{nodelayer}
    \node [style=\commandkey{style1}] (0) at (0, 1.1) {$#1$};
    \node [style=\commandkey{style2}] (1) at (0, -0.5) {$#2$};
    \node [style=none] (2) at (0, -1.75) {};
  \end{pgfonlayer}
  \begin{pgfonlayer}{edgelayer}
    \draw (0) to (1);
    \draw (1) to (2.center);
  \end{pgfonlayer}
\end{tikzpicture}%
\,}

\newkeycommand{\dkpointmap}[style1=dkpoint,style2=dmap][2]{\,%
\begin{tikzpicture}
  \begin{pgfonlayer}{nodelayer}
    \node [style=\commandkey{style1}] (0) at (0, -1.2) {$#1$};
    \node [style=\commandkey{style2}] (1) at (0, 0.4) {$#2$};
    \node [style=none] (2) at (0, 1.7) {};
  \end{pgfonlayer}
  \begin{pgfonlayer}{edgelayer}
    \draw[style=boldedge] (0) to (1);
    \draw[style=boldedge] (1) to (2.center);
  \end{pgfonlayer}
\end{tikzpicture}%
\,}

\newkeycommand{\dkpointmapx}[style1=dkpoint,style2=dmap][2]{\,%
\begin{tikzpicture}
  \begin{pgfonlayer}{nodelayer}
    \node [style=\commandkey{style1}] (0) at (0, -1.4) {$#1$};
    \node [style=\commandkey{style2}] (1) at (0, 0.6) {$#2$};
    \node [style=none] (2) at (0, 1.9) {};
  \end{pgfonlayer}
  \begin{pgfonlayer}{edgelayer}
    \draw[style=boldedge] (0) to (1);
    \draw[style=boldedge] (1) to (2.center);
  \end{pgfonlayer}
\end{tikzpicture}%
\,}

\newkeycommand{\boxcopointmapdag}[style1=map,style2=copoint][2]{\,%
\begin{tikzpicture}
  \begin{pgfonlayer}{nodelayer}
    \node [style=none] (0) at (0, -1.75) {};
    \node [style=\commandkey{style1}] (1) at (0, -0.5) {$#1$};
    \node [style=\commandkey{style2}] (2) at (0, 1) {$#2$};
  \end{pgfonlayer}
  \begin{pgfonlayer}{edgelayer}
    \draw (0.center) to (1);
    \draw (1) to (2);
  \end{pgfonlayer}
\end{tikzpicture}%
\,}

\newkeycommand{\kpointdagmap}[style1=map,style2=kpointdag][2]{\,%
\begin{tikzpicture}
  \begin{pgfonlayer}{nodelayer}
    \node [style=none] (0) at (0, -1.75) {};
    \node [style=\commandkey{style1}] (1) at (0, -0.5) {$#1$};
    \node [style=\commandkey{style2}] (2) at (0, 1.1) {$#2$};
  \end{pgfonlayer}
  \begin{pgfonlayer}{edgelayer}
    \draw (0.center) to (1);
    \draw (1) to (2);
  \end{pgfonlayer}
\end{tikzpicture}%
\,}

\newkeycommand{\sandwichmap}[style1=point,style2=map,style3=copoint][3]{\,%
\begin{tikzpicture}
  \begin{pgfonlayer}{nodelayer}
    \node [style=\commandkey{style1}] (0) at (0, -1.50) {$#1$};
    \node [style=\commandkey{style2}] (1) at (0, 0) {$#2$};
    \node [style=\commandkey{style3}] (2) at (0, 1.50) {$#3$};
  \end{pgfonlayer}
  \begin{pgfonlayer}{edgelayer}
    \draw (0) to (1);
    \draw (1) to (2);
  \end{pgfonlayer}
\end{tikzpicture}%
}

\newkeycommand{\kpointsandwichmap}[style1=kpoint,style2=map,style3=kpointdag][3]{\,%
\begin{tikzpicture}
  \begin{pgfonlayer}{nodelayer}
    \node [style=\commandkey{style1}] (0) at (0, -1.5) {$#1$};
    \node [style=\commandkey{style2}] (1) at (0, 0) {$#2$};
    \node [style=\commandkey{style3}] (2) at (0, 1.5) {$#3$};
  \end{pgfonlayer}
  \begin{pgfonlayer}{edgelayer}
    \draw (0) to (1);
    \draw (1) to (2);
  \end{pgfonlayer}
\end{tikzpicture}%
}

\newkeycommand{\sandwichmapconj}[style1=point,style2=mapconj,style3=copoint][3]{\,% 
\begin{tikzpicture}
  \begin{pgfonlayer}{nodelayer}
    \node [style=\commandkey{style1}] (0) at (0, -1.50) {$#1$};
    \node [style=\commandkey{style2}] (1) at (0, 0) {$#2$};
    \node [style=\commandkey{style3}] (2) at (0, 1.50) {$#3$};
  \end{pgfonlayer}
  \begin{pgfonlayer}{edgelayer}
    \draw (0) to (1);
    \draw (1) to (2);
  \end{pgfonlayer}
\end{tikzpicture}%
}

\newkeycommand{\sandwichtwo}[style1=point,style2=map,style3=map,style4=copoint][4]{\,%
\begin{tikzpicture}
  \begin{pgfonlayer}{nodelayer}
    \node [style=\commandkey{style2}] (0) at (0, -1) {$#2$};
    \node [style=\commandkey{style3}] (1) at (0, 1) {$#3$};
    \node [style=\commandkey{style1}] (2) at (0, -2.6) {$#1$};
    \node [style=\commandkey{style4}] (3) at (0, 2.6) {$#4$};
  \end{pgfonlayer}
  \begin{pgfonlayer}{edgelayer}
    \draw (2) to (0);
    \draw (0) to (1);
    \draw (1) to (3);
  \end{pgfonlayer}
\end{tikzpicture}\,}

\newkeycommand{\longbraket}[style1=point,style2=copoint][2]{\,% 
\begin{tikzpicture}
  \begin{pgfonlayer}{nodelayer}
    \node [style=\commandkey{style1}] (0) at (0, -2.6) {$#1$};
    \node [style=\commandkey{style2}] (1) at (0, 2.6) {$#2$};
  \end{pgfonlayer}
  \begin{pgfonlayer}{edgelayer}
    \draw (0) to (1);
  \end{pgfonlayer}
\end{tikzpicture}\,}

\newkeycommand{\onb}[style=point]{\ensuremath{\{\,\tikz{\node[style=\commandkey{style}] (x) at (0,-0.3) {$j$};\draw (x)--(0,0.7);}\,\}}\xspace}

\newkeycommand{\phase}[style=white phase dot][1]{\,\begin{tikzpicture}
    \begin{pgfonlayer}{nodelayer}
        \node [style=none] (0) at (0, 1) {};
        \node [style=\commandkey{style}] (2) at (0, -0) {$#1$}; 
        \node [style=none] (3) at (0, -1) {};
    \end{pgfonlayer}
    \begin{pgfonlayer}{edgelayer}
        \draw (2) to (0.center);
        \draw (3.center) to (2);
    \end{pgfonlayer}
\end{tikzpicture}\,}

\newkeycommand{\dphase}[style=white phase ddot][1]{\,\begin{tikzpicture}
    \begin{pgfonlayer}{nodelayer}
        \node [style=none] (0) at (0, 1.25) {};
        \node [style=\commandkey{style}] (2) at (0, -0) {$#1$};
        \node [style=none] (3) at (0, -1.25) {};
    \end{pgfonlayer}
    \begin{pgfonlayer}{edgelayer}
        \draw [boldedge] (2) to (0.center);
        \draw [boldedge] (3.center) to (2);
    \end{pgfonlayer}
\end{tikzpicture}\,}

\newkeycommand{\dphasepoint}[style=white phase ddot][1]{\,\begin{tikzpicture}[yshift=-3mm]
  \begin{pgfonlayer}{nodelayer}
    \node [style=none] (0) at (0, 1) {};
    \node [style=\commandkey{style}] (1) at (0, 0) {$#1$};
  \end{pgfonlayer}
  \begin{pgfonlayer}{edgelayer}
    \draw [style=boldedge] (0.center) to (1);
  \end{pgfonlayer}
\end{tikzpicture}\,}

\newkeycommand{\dphasepointgray}[style=gray phase ddot][1]{\,\begin{tikzpicture}[yshift=-3mm]
  \begin{pgfonlayer}{nodelayer}
    \node [style=none] (0) at (0, 1) {};
    \node [style=\commandkey{style}] (1) at (0, 0) {$#1$};
  \end{pgfonlayer}
  \begin{pgfonlayer}{edgelayer}
    \draw [style=boldedge] (0.center) to (1);
  \end{pgfonlayer}
\end{tikzpicture}\,}

\newkeycommand{\dphasecopoint}[style=white phase ddot][1]{\,\begin{tikzpicture}[yshift=3mm,yscale=-1]
  \begin{pgfonlayer}{nodelayer}
    \node [style=none] (0) at (0, 1) {};
    \node [style=\commandkey{style}] (1) at (0, 0) {$#1$};
  \end{pgfonlayer}
  \begin{pgfonlayer}{edgelayer}
    \draw [style=boldedge] (0.center) to (1);
  \end{pgfonlayer}
\end{tikzpicture}\,}

\newkeycommand{\dphasepointsm}[style=white phase ddot][1]{\,\begin{tikzpicture}[yshift=-3mm]
  \begin{pgfonlayer}{nodelayer}
    \node [style=none] (0) at (0, 1) {};
    \node [style=\commandkey{style}] (1) at (0, 0) {\footnotesize\!$#1$\!};
  \end{pgfonlayer}
  \begin{pgfonlayer}{edgelayer}
    \draw [style=boldedge] (0.center) to (1);
  \end{pgfonlayer}
\end{tikzpicture}\,}

\newkeycommand{\phasepoint}[style=white phase dot][1]{\,\begin{tikzpicture}[yshift=-3mm]
  \begin{pgfonlayer}{nodelayer}
    \node [style=none] (0) at (0, 1) {};
    \node [style=\commandkey{style}] (1) at (0, 0) {$#1$};
  \end{pgfonlayer}
  \begin{pgfonlayer}{edgelayer}
    \draw (0.center) to (1);
  \end{pgfonlayer}
\end{tikzpicture}\,}

\newkeycommand{\phasecopoint}[style=white phase dot][1]{\,\begin{tikzpicture}[yshift=3mm]
  \begin{pgfonlayer}{nodelayer}
    \node [style=none] (0) at (0, -1) {};
    \node [style=\commandkey{style}] (1) at (0, 0) {$#1$};
  \end{pgfonlayer}
  \begin{pgfonlayer}{edgelayer}
    \draw (0.center) to (1);
  \end{pgfonlayer}
\end{tikzpicture}\,}

\newkeycommand{\kpointbraket}[style1=kpoint,style2=kpoint adjoint,estyle=][2]{\,%
\begin{tikzpicture}
  \begin{pgfonlayer}{nodelayer}
    \node [style=\commandkey{style1}] (0) at (0, -0.735) {$#1$};
    \node [style=\commandkey{style2}] (1) at (0, 0.735) {$#2$};
  \end{pgfonlayer}
  \begin{pgfonlayer}{edgelayer}
    \draw [\commandkey{estyle}] (0) to (1);
  \end{pgfonlayer}
\end{tikzpicture}\,}

\newkeycommand{\kkpointbraket}[style1=kpoint,style2=kpoint adjoint,estyle=][2]{\,%
\begin{tikzpicture}
  \begin{pgfonlayer}{nodelayer}
    \node [style=\commandkey{style1}] (0) at (0, -0.685) {$#1$};
    \node [style=\commandkey{style2}] (1) at (0, 0.685) {$#2$};
  \end{pgfonlayer}
  \begin{pgfonlayer}{edgelayer}
    \draw [\commandkey{estyle}] (0) to (1);
  \end{pgfonlayer}
\end{tikzpicture}\,}

\newkeycommand{\kpointbraketx}[style1=kpoint,style2=kpoint adjoint,estyle=][2]{\,%
\begin{tikzpicture}
  \begin{pgfonlayer}{nodelayer}
    \node [style=\commandkey{style1}] (0) at (0, -0.885) {$#1$};
    \node [style=\commandkey{style2}] (1) at (0, 0.885) {$#2$};
  \end{pgfonlayer}
  \begin{pgfonlayer}{edgelayer}
    \draw [\commandkey{estyle}] (0) to (1);
  \end{pgfonlayer}
\end{tikzpicture}\,}

\tikzstyle{cdiag}=[matrix of math nodes, row sep=3em, column sep=3em, text height=1.5ex, text depth=0.25ex,inner sep=0.5em]
\tikzstyle{arrow above}=[transform canvas={yshift=0.5ex}]
\tikzstyle{arrow below}=[transform canvas={yshift=-0.5ex}]

\usepackage{tikz}
\usetikzlibrary{decorations.markings}
\usetikzlibrary{shapes.geometric}

\pgfdeclarelayer{edgelayer}
\pgfdeclarelayer{nodelayer}
\pgfsetlayers{edgelayer,nodelayer,main}

\tikzstyle{none}=[inner sep=0pt]
\definecolor{hexcolor0xff0000}{rgb}{1.000,0.000,0.000}
\definecolor{hexcolor0x000000}{rgb}{0.000,0.000,0.000}
\definecolor{hexcolor0x00ff00}{rgb}{0.000,1.000,0.000}
\definecolor{hexcolor0x000000}{rgb}{0.000,0.000,0.000}
\definecolor{hexcolor0xffff00}{rgb}{1.000,1.000,0.000}
\definecolor{hexcolor0xffffff}{rgb}{1.000,1.000,1.000}

\tikzstyle{rn}=[circle,fill=hexcolor0xff0000,draw=hexcolor0x000000,line width=0.8 pt]
\tikzstyle{gn}=[circle,fill=hexcolor0x00ff00,draw=hexcolor0x000000,line width=0.8 pt]
\tikzstyle{yn}=[circle,fill=hexcolor0xffff00,draw=hexcolor0x000000,line width=0.8 pt]
\tikzstyle{wn}=[circle,fill=hexcolor0xffffff,draw=hexcolor0x000000,line width=0.8 pt]
\tikzstyle{wnthick}=[circle,fill=hexcolor0xffffff,draw=hexcolor0x000000,line width=2.500]

\tikzstyle{simple}=[-,draw=hexcolor0x000000,line width=2.000]
\tikzstyle{arrow}=[-,draw=hexcolor0x000000,postaction={decorate},decoration={markings,mark=at position .5 with {\arrow{>}}},line width=2.000]
\tikzstyle{tick}=[-,draw=hexcolor0x000000,postaction={decorate},decoration={markings,mark=at position .5 with {\draw (0,-0.1) -- (0,0.1);}},line width=2.000]
\tikzstyle{halfthickness}=[-,draw=hexcolor0x000000,line width=0.500]
\tikzstyle{thick}=[-,draw=hexcolor0x000000,line width=2.500]
\tikzstyle{thicker}=[-,draw=hexcolor0x000000,line width=4.000]

\tikzstyle{env}=[copoint,regular polygon rotate=0,minimum width=0.2cm, fill=black]

\tikzstyle{probs}=[shape=semicircle,fill=white,draw=black,shape border rotate=180,minimum width=1.2cm]

\tikzstyle{every picture}=[baseline=-0.25em,scale=0.5]
\tikzstyle{dotpic}=[] % for backwards-compatibility
\tikzstyle{diredges}=[every to/.style={diredge}]
\tikzstyle{math matrix}=[matrix of math nodes,left delimiter=(,right delimiter=),inner sep=2pt,column sep=1em,row sep=0.5em,nodes={inner sep=0pt},text height=1.5ex, text depth=0.25ex]

\tikzstyle{inline text}=[text height=1.5ex, text depth=0.25ex,yshift=0.5mm]
\tikzstyle{label}=[font=\footnotesize,text height=1.5ex, text depth=0.25ex,yshift=0.5mm]
\tikzstyle{left label}=[label,anchor=east,xshift=1.5mm]
\tikzstyle{right label}=[label,anchor=west,xshift=-1.5mm]

\tikzstyle{braceedge}=[decorate,decoration={brace,amplitude=2mm,raise=-1mm}]
\tikzstyle{small braceedge}=[decorate,decoration={brace,amplitude=1mm,raise=-1mm}]

\tikzstyle{doubled}=[line width=1.6pt] % set the line width for all doubled (quantum) maps/wires
\tikzstyle{boldedge}=[doubled,shorten <=-0.17mm,shorten >=-0.17mm]
\tikzstyle{boldedgegray}=[doubled,gray,shorten <=-0.17mm,shorten >=-0.17mm]

\tikzstyle{semidoubled}=[line width=1.4pt] % set the line width for all doubled (quantum) maps/wires
\tikzstyle{semiboldedgegray}=[semidoubled,gray,shorten <=-0.17mm,shorten >=-0.17mm]

\tikzstyle{boldedgedashed}=[very thick,dashed,shorten <=-0.17mm,shorten >=-0.17mm]
\tikzstyle{vboldedgedashed}=[doubled,dashed,shorten <=-0.17mm,shorten >=-0.17mm]
\tikzstyle{left hook arrow}=[left hook-latex]
\tikzstyle{right hook arrow}=[right hook-latex]
\tikzstyle{sembracket}=[line width=0.5pt,shorten <=-0.07mm,shorten >=-0.07mm]

\tikzstyle{causal edge}=[->,thick,gray]
\tikzstyle{causal nondir}=[thick,gray]
\tikzstyle{timeline}=[thick,gray, dashed]

\tikzstyle{cedge}=[<->,thick,gray!70!white]

\tikzstyle{empty diagram}=[draw=gray!40!white,dashed,shape=rectangle,minimum width=1cm,minimum height=1cm]
\tikzstyle{empty diagram small}=[draw=gray!50!white,dashed,shape=rectangle,minimum width=0.6cm,minimum height=0.5cm]

\tikzstyle{dot}=[inner sep=0mm,minimum width=2mm,minimum height=2mm,draw,shape=circle]
\tikzstyle{ddot}=[inner sep=0mm, doubled, minimum width=2.5mm,minimum height=2.5mm,draw,shape=circle]

\tikzstyle{black dot}=[dot,fill=black]
\tikzstyle{white dot}=[dot,fill=white,,text depth=-0.2mm]
\tikzstyle{green dot}=[white dot] % for backwards-compatibility
\tikzstyle{gray dot}=[dot,fill=gray!40!white,,text depth=-0.2mm]
\tikzstyle{red dot}=[gray dot] % for backwards-compatibility

\tikzstyle{black ddot}=[ddot,fill=black]
\tikzstyle{white ddot}=[ddot,fill=white]
\tikzstyle{gray ddot}=[ddot,fill=gray!40!white]

\tikzstyle{gray edge}=[gray!40!white]

\tikzstyle{small dot}=[inner sep=0.3mm,minimum width=0pt,minimum height=0pt,draw,shape=circle]

\tikzstyle{small black dot}=[small dot,fill=black]
\tikzstyle{small white dot}=[small dot,fill=white]
\tikzstyle{small gray dot}=[small dot,fill=gray!40!white]

\tikzstyle{causal dot}=[inner sep=0.4mm,minimum width=0pt,minimum height=0pt,draw=white,shape=circle,fill=gray!40!white]

\tikzstyle{phase dimensions}=[minimum size=5mm,font=\footnotesize,rectangle,rounded corners=2.5mm,inner sep=0.2mm,outer sep=-2mm]
\tikzstyle{dphase dimensions}=[minimum size=5mm,font=\footnotesize,rectangle,rounded corners=2.5mm,inner sep=0.2mm,outer sep=-2mm]
\tikzstyle{white phase dot}=[dot,fill=white,phase dimensions]
\tikzstyle{white phase ddot}=[ddot,fill=white,dphase dimensions]
\tikzstyle{green phase ddot}=[ddot,fill=green,dphase dimensions]

\tikzstyle{white rect ddot}=[draw=black,fill=white,doubled,minimum size=5mm,font=\footnotesize,rectangle,rounded corners=2.5mm,inner sep=0.2mm]
\tikzstyle{gray rect ddot}=[draw=black,fill=gray!40!white,doubled,minimum size=6mm,font=\footnotesize,rectangle,rounded corners=3mm]

\tikzstyle{gray phase dot}=[dot,fill=gray!40!white,phase dimensions]
\tikzstyle{gray phase ddot}=[ddot,fill=gray!40!white,dphase dimensions]
\tikzstyle{red phase ddot}=[ddot,fill=red,dphase dimensions]

\tikzstyle{grey phase dot}=[gray phase dot]
\tikzstyle{grey phase ddot}=[gray phase ddot]

\tikzstyle{small phase dimensions}=[minimum size=4mm,font=\tiny,rectangle,rounded corners=2mm,inner sep=0.2mm,outer sep=-2mm]
\tikzstyle{small dphase dimensions}=[minimum size=4mm,font=\tiny,rectangle,rounded corners=2mm,inner sep=0.2mm,outer sep=-2mm]

\tikzstyle{small gray phase dot}=[dot,fill=gray!40!white,small phase dimensions]
\tikzstyle{small gray phase ddot}=[ddot,fill=gray!40!white,small dphase dimensions]

\tikzstyle{small map}=[draw,shape=rectangle,minimum height=4mm,minimum width=4mm,fill=white]

\tikzstyle{cnot}=[fill=white,shape=circle,inner sep=-1.4pt]

\tikzstyle{asym hadamard}=[fill=white,draw,shape=NEbox,inner sep=0.6mm,font=\footnotesize,minimum height=4mm]
\tikzstyle{asym hadamard conj}=[fill=white,draw,shape=NWbox,inner sep=0.6mm,font=\footnotesize,minimum height=4mm]
\tikzstyle{asym hadamard dag}=[fill=white,draw,shape=SEbox,inner sep=0.6mm,font=\footnotesize,minimum height=4mm]

\tikzstyle{hadamard}=[fill=white,draw,inner sep=0.6mm,font=\footnotesize,minimum height=4mm,minimum width=4mm]
\tikzstyle{small hadamard}=[fill=white,draw,inner sep=0.6mm,minimum height=1.5mm,minimum width=1.5mm]
\tikzstyle{dhadamard}=[hadamard,doubled]
\tikzstyle{small dhadamard}=[small hadamard,doubled]
\tikzstyle{small dhadamard rotate}=[small hadamard,doubled,rotate=45]
\tikzstyle{antipode}=[white dot,inner sep=0.3mm,font=\footnotesize]

\tikzstyle{scalar}=[diamond,draw,inner sep=0.5pt,font=\small]
\tikzstyle{dscalar}=[diamond,doubled, draw,inner sep=0.5pt,font=\small]

\tikzstyle{small box}=[rectangle,inline text,fill=white,draw,minimum height=5mm,yshift=-0.5mm,minimum width=5mm,font=\small]
\tikzstyle{small gray box}=[small box,fill=gray!30]
\tikzstyle{medium box}=[rectangle,inline text,fill=white,draw,minimum height=5mm,yshift=-0.5mm,minimum width=10mm,font=\small]
\tikzstyle{square box}=[small box] % for backwards-compatibility
\tikzstyle{medium gray box}=[small box,fill=gray!30]
\tikzstyle{semilarge box}=[rectangle,inline text,fill=white,draw,minimum height=5mm,yshift=-0.5mm,minimum width=12.5mm,font=\small]
\tikzstyle{large box}=[rectangle,inline text,fill=white,draw,minimum height=5mm,yshift=-0.5mm,minimum width=15mm,font=\small]
\tikzstyle{large gray box}=[small box,fill=gray!30]

\tikzstyle{Bayes box}=[rectangle,fill=black,draw, minimum height=3mm, minimum width=3mm]

\tikzstyle{gray square point}=[small box,fill=gray!50]

\tikzstyle{dphase box white}=[dhadamard]
\tikzstyle{dphase box gray}=[dhadamard,fill=gray!50!white]

\tikzstyle{point}=[regular polygon,regular polygon sides=3,draw,scale=0.75,inner sep=-0.5pt,minimum width=9mm,fill=white,regular polygon rotate=180]
\tikzstyle{copoint}=[regular polygon,regular polygon sides=3,draw,scale=0.75,inner sep=-0.5pt,minimum width=9mm,fill=white]
\tikzstyle{dpoint}=[point,doubled]
\tikzstyle{dcopoint}=[copoint,doubled]

\tikzstyle{wide copoint}=[fill=white,draw,shape=isosceles triangle,shape border rotate=90,isosceles triangle stretches=true,inner sep=0pt,minimum width=1.5cm,minimum height=6.12mm]
\tikzstyle{wide point}=[fill=white,draw,shape=isosceles triangle,shape border rotate=-90,isosceles triangle stretches=true,inner sep=0pt,minimum width=1.5cm,minimum height=6.12mm,yshift=-0.0mm]
\tikzstyle{wide point plus}=[fill=white,draw,shape=isosceles triangle,shape border rotate=-90,isosceles triangle stretches=true,inner sep=0pt,minimum width=1.74cm,minimum height=7mm,yshift=-0.0mm]

\tikzstyle{wide dpoint}=[fill=white,doubled,draw,shape=isosceles triangle,shape border rotate=-90,isosceles triangle stretches=true,inner sep=0pt,minimum width=1.5cm,minimum height=6.12mm,yshift=-0.0mm]
\tikzstyle{wide dcopoint}=[fill=white,doubled,draw,shape=isosceles triangle,shape border rotate=90,isosceles triangle stretches=true,inner sep=0pt,minimum width=1.5cm,minimum height=6.12mm,yshift=-0.0mm]

\tikzstyle{tinypoint}=[regular polygon,regular polygon sides=3,draw,scale=0.55,inner sep=-0.15pt,minimum width=6mm,fill=white,regular polygon rotate=180]

\tikzstyle{white point}=[point]
\tikzstyle{white dpoint}=[dpoint]
\tikzstyle{green point}=[white point] % for backwards-compatibility
\tikzstyle{white copoint}=[copoint]
\tikzstyle{gray point}=[point,fill=gray!40!white]
\tikzstyle{gray dpoint}=[gray point,doubled]
\tikzstyle{red point}=[gray point] % for backwards-compatibility
\tikzstyle{gray copoint}=[copoint,fill=gray!40!white]
\tikzstyle{gray dcopoint}=[gray copoint,doubled]

\tikzstyle{white point guide}=[regular polygon,regular polygon sides=3,font=\scriptsize,draw,scale=0.65,inner sep=-0.5pt,minimum width=9mm,fill=white,regular polygon rotate=180]

\tikzstyle{black point}=[point,fill=black,font=\color{white}]
\tikzstyle{black copoint}=[copoint,fill=black,font=\color{white}]

\tikzstyle{tiny gray point}=[tinypoint,fill=gray!40!white]

\tikzstyle{diredge}=[->]
\tikzstyle{ddiredge}=[<->]
\tikzstyle{rdiredge}=[<-]
\tikzstyle{thickdiredge}=[->, very thick]
\tikzstyle{pointer edge}=[->,very thick,gray]
\tikzstyle{pointer edge part}=[very thick,gray]
\tikzstyle{dashed edge}=[dashed]
\tikzstyle{thick dashed edge}=[very thick,dashed]
\tikzstyle{thick gray dashed edge}=[thick dashed edge,gray!40]
\tikzstyle{thick map edge}=[very thick,|->]

\makeatletter
\newcommand{\boxshape}[3]{%
\pgfdeclareshape{#1}{
\inheritsavedanchors[from=rectangle] % this is nearly a rectangle
\inheritanchorborder[from=rectangle]
\inheritanchor[from=rectangle]{center}
\inheritanchor[from=rectangle]{north}
\inheritanchor[from=rectangle]{south}
\inheritanchor[from=rectangle]{west}
\inheritanchor[from=rectangle]{east}
\backgroundpath{% this is new
\southwest \pgf@xa=\pgf@x \pgf@ya=\pgf@y
\northeast \pgf@xb=\pgf@x \pgf@yb=\pgf@y

\@tempdima=#2
\@tempdimb=#3

\pgfpathmoveto{\pgfpoint{\pgf@xa - 5pt + \@tempdima}{\pgf@ya}}
\pgfpathlineto{\pgfpoint{\pgf@xa - 5pt - \@tempdima}{\pgf@yb}}
\pgfpathlineto{\pgfpoint{\pgf@xb + 5pt + \@tempdimb}{\pgf@yb}}
\pgfpathlineto{\pgfpoint{\pgf@xb + 5pt - \@tempdimb}{\pgf@ya}}
\pgfpathlineto{\pgfpoint{\pgf@xa - 5pt + \@tempdima}{\pgf@ya}}
\pgfpathclose
}
}}

\boxshape{NEbox}{0pt}{5pt}
\boxshape{SEbox}{0pt}{-5pt}
\boxshape{NWbox}{5pt}{0pt}
\boxshape{SWbox}{-5pt}{0pt}
\boxshape{EBox}{-3pt}{3pt}
\boxshape{WBox}{3pt}{-3pt}
\makeatother

\tikzstyle{cloud}=[shape=cloud,draw,minimum width=1.5cm,minimum height=1.5cm]

\tikzstyle{map}=[draw,shape=NEbox,inner sep=2pt,minimum height=6mm,fill=white]
\tikzstyle{dashedmap}=[draw,dashed,shape=NEbox,inner sep=2pt,minimum height=6mm,fill=white]
\tikzstyle{mapdag}=[draw,shape=SEbox,inner sep=2pt,minimum height=6mm,fill=white]
\tikzstyle{mapadj}=[draw,shape=SEbox,inner sep=2pt,minimum height=6mm,fill=white]
\tikzstyle{maptrans}=[draw,shape=SWbox,inner sep=2pt,minimum height=6mm,fill=white]
\tikzstyle{mapconj}=[draw,shape=NWbox,inner sep=2pt,minimum height=6mm,fill=white]

\tikzstyle{langmap}=[draw,shape=NEbox,inner sep=2pt,minimum height=2.4mm,minimum width=3.2mm,fill=white]
\tikzstyle{langmaptrans}=[draw,shape=SWbox,inner sep=2pt,minimum height=2.4mm,minimum width=3.2mm,fill=white]

\tikzstyle{medium map}=[draw,shape=NEbox,inner sep=2pt,minimum height=6mm,fill=white,minimum width=7mm]
\tikzstyle{medium map dag}=[draw,shape=SEbox,inner sep=2pt,minimum height=6mm,fill=white,minimum width=7mm]
\tikzstyle{medium map adj}=[draw,shape=SEbox,inner sep=2pt,minimum height=6mm,fill=white,minimum width=7mm]
\tikzstyle{medium map trans}=[draw,shape=SWbox,inner sep=2pt,minimum height=6mm,fill=white,minimum width=7mm]
\tikzstyle{medium map conj}=[draw,shape=NWbox,inner sep=2pt,minimum height=6mm,fill=white,minimum width=7mm]
\tikzstyle{semilarge map}=[draw,shape=NEbox,inner sep=2pt,minimum height=6mm,fill=white,minimum width=9.5mm]
\tikzstyle{semilarge map trans}=[draw,shape=SWbox,inner sep=2pt,minimum height=6mm,fill=white,minimum width=9.5mm]
\tikzstyle{semilarge map adj}=[draw,shape=SEbox,inner sep=2pt,minimum height=6mm,fill=white,minimum width=9.5mm]
\tikzstyle{semilarge map dag}=[draw,shape=SEbox,inner sep=2pt,minimum height=6mm,fill=white,minimum width=9.5mm]
\tikzstyle{semilarge map conj}=[draw,shape=NWbox,inner sep=2pt,minimum height=6mm,fill=white,minimum width=9.5mm]
\tikzstyle{large map}=[draw,shape=NEbox,inner sep=2pt,minimum height=6mm,fill=white,minimum width=12mm]
\tikzstyle{large map conj}=[draw,shape=NWbox,inner sep=2pt,minimum height=6mm,fill=white,minimum width=12mm]
\tikzstyle{very large map}=[draw,shape=NEbox,inner sep=2pt,minimum height=6mm,fill=white,minimum width=17mm]

\tikzstyle{medium dmap}=[draw,doubled,shape=NEbox,inner sep=2pt,minimum height=6mm,fill=white,minimum width=7mm]
\tikzstyle{medium dmap dag}=[draw,doubled,shape=SEbox,inner sep=2pt,minimum height=6mm,fill=white,minimum width=7mm]
\tikzstyle{medium dmap adj}=[draw,doubled,shape=SEbox,inner sep=2pt,minimum height=6mm,fill=white,minimum width=7mm]
\tikzstyle{medium dmap trans}=[draw,doubled,shape=SWbox,inner sep=2pt,minimum height=6mm,fill=white,minimum width=7mm]
\tikzstyle{medium dmap conj}=[draw,doubled,shape=NWbox,inner sep=2pt,minimum height=6mm,fill=white,minimum width=7mm]
\tikzstyle{semilarge dmap}=[draw,doubled,shape=NEbox,inner sep=2pt,minimum height=6mm,fill=white,minimum width=9.5mm]
\tikzstyle{semilarge dmap trans}=[draw,doubled,shape=SWbox,inner sep=2pt,minimum height=6mm,fill=white,minimum width=9.5mm]
\tikzstyle{semilarge dmap adj}=[draw,doubled,shape=SEbox,inner sep=2pt,minimum height=6mm,fill=white,minimum width=9.5mm]
\tikzstyle{semilarge dmap dag}=[draw,doubled,shape=SEbox,inner sep=2pt,minimum height=6mm,fill=white,minimum width=9.5mm]
\tikzstyle{semilarge dmap conj}=[draw,doubled,shape=NWbox,inner sep=2pt,minimum height=6mm,fill=white,minimum width=9.5mm]
\tikzstyle{large dmap}=[draw,doubled,shape=NEbox,inner sep=2pt,minimum height=6mm,fill=white,minimum width=12mm]
\tikzstyle{large dmap conj}=[draw,doubled,shape=NWbox,inner sep=2pt,minimum height=6mm,fill=white,minimum width=12mm]
\tikzstyle{large dmap trans}=[draw,doubled,shape=SWbox,inner sep=2pt,minimum height=6mm,fill=white,minimum width=12mm]
\tikzstyle{large dmap adj}=[draw,doubled,shape=SEbox,inner sep=2pt,minimum height=6mm,fill=white,minimum width=12mm]
\tikzstyle{large dmap dag}=[draw,doubled,shape=SEbox,inner sep=2pt,minimum height=6mm,fill=white,minimum width=12mm]
\tikzstyle{very large dmap}=[draw,doubled,shape=NEbox,inner sep=2pt,minimum height=6mm,fill=white,minimum width=19.5mm]

\tikzstyle{muxbox}=[draw,shape=rectangle,minimum height=3mm,minimum width=3mm,fill=white]
\tikzstyle{dmuxbox}=[muxbox,doubled]

\tikzstyle{box}=[draw,shape=rectangle,inner sep=2pt,minimum height=6mm,minimum width=6mm,fill=white]
\tikzstyle{dbox}=[draw,doubled,shape=rectangle,inner sep=2pt,minimum height=6mm,minimum width=6mm,fill=white]
\tikzstyle{dmap}=[draw,doubled,shape=NEbox,inner sep=2pt,minimum height=6mm,fill=white]
\tikzstyle{dmapdag}=[draw,doubled,shape=SEbox,inner sep=2pt,minimum height=6mm,fill=white]
\tikzstyle{dmapadj}=[draw,doubled,shape=SEbox,inner sep=2pt,minimum height=6mm,fill=white]
\tikzstyle{dmaptrans}=[draw,doubled,shape=SWbox,inner sep=2pt,minimum height=6mm,fill=white]
\tikzstyle{dmapconj}=[draw,doubled,shape=NWbox,inner sep=2pt,minimum height=6mm,fill=white]

\tikzstyle{ddmap}=[draw,doubled,dashed,shape=NEbox,inner sep=2pt,minimum height=6mm,fill=white]
\tikzstyle{ddmapdag}=[draw,doubled,dashed,shape=SEbox,inner sep=2pt,minimum height=6mm,fill=white]
\tikzstyle{ddmapadj}=[draw,doubled,dashed,shape=SEbox,inner sep=2pt,minimum height=6mm,fill=white]
\tikzstyle{ddmaptrans}=[draw,doubled,dashed,shape=SWbox,inner sep=2pt,minimum height=6mm,fill=white]
\tikzstyle{ddmapconj}=[draw,doubled,dashed,shape=NWbox,inner sep=2pt,minimum height=6mm,fill=white]

\boxshape{sNEbox}{0pt}{3pt}
\boxshape{sSEbox}{0pt}{-3pt}
\boxshape{sNWbox}{3pt}{0pt}
\boxshape{sSWbox}{-3pt}{0pt}
\tikzstyle{smap}=[draw,shape=sNEbox,fill=white]
\tikzstyle{smapdag}=[draw,shape=sSEbox,fill=white]
\tikzstyle{smapadj}=[draw,shape=sSEbox,fill=white]
\tikzstyle{smaptrans}=[draw,shape=sSWbox,fill=white]
\tikzstyle{smapconj}=[draw,shape=sNWbox,fill=white]

\tikzstyle{dsmap}=[draw,dashed,shape=sNEbox,fill=white]
\tikzstyle{dsmapdag}=[draw,dashed,shape=sSEbox,fill=white]
\tikzstyle{dsmaptrans}=[draw,dashed,shape=sSWbox,fill=white]
\tikzstyle{dsmapconj}=[draw,dashed,shape=sNWbox,fill=white]

\boxshape{mNEbox}{0pt}{10pt}
\boxshape{mSEbox}{0pt}{-10pt}
\boxshape{mNWbox}{10pt}{0pt}
\boxshape{mSWbox}{-10pt}{0pt}
\tikzstyle{mmap}=[draw,shape=mNEbox]
\tikzstyle{mmapdag}=[draw,shape=mSEbox]
\tikzstyle{mmaptrans}=[draw,shape=mSWbox]
\tikzstyle{mmapconj}=[draw,shape=mNWbox]

\tikzstyle{mmapgray}=[draw,fill=gray!40!white,shape=mNEbox]
\tikzstyle{smapgray}=[draw,fill=gray!40!white,shape=sNEbox]

\makeatletter
\pgfdeclareshape{cornerpoint}{
\inheritsavedanchors[from=rectangle] % this is nearly a rectangle
\inheritanchorborder[from=rectangle]
\inheritanchor[from=rectangle]{center}
\inheritanchor[from=rectangle]{north}
\inheritanchor[from=rectangle]{south}
\inheritanchor[from=rectangle]{west}
\inheritanchor[from=rectangle]{east}
\backgroundpath{% this is new
\southwest \pgf@xa=\pgf@x \pgf@ya=\pgf@y
\northeast \pgf@xb=\pgf@x \pgf@yb=\pgf@y

\pgfmathsetmacro{\pgf@shorten@left}{\pgfkeysvalueof{/tikz/shorten left}}
\pgfmathsetmacro{\pgf@shorten@right}{\pgfkeysvalueof{/tikz/shorten right}}

\pgfpathmoveto{\pgfpoint{0.5 * (\pgf@xa + \pgf@xb)}{\pgf@ya - 5pt}}
\pgfpathlineto{\pgfpoint{\pgf@xa - 8pt + \pgf@shorten@left}{\pgf@yb - 1.5 * \pgf@shorten@left}}
\pgfpathlineto{\pgfpoint{\pgf@xa - 8pt + \pgf@shorten@left}{\pgf@yb}}
\pgfpathlineto{\pgfpoint{\pgf@xb + 8pt - \pgf@shorten@right}{\pgf@yb}}
\pgfpathlineto{\pgfpoint{\pgf@xb + 8pt - \pgf@shorten@right}{\pgf@yb - 1.5 * \pgf@shorten@right}}
\pgfpathclose
}
}

\pgfdeclareshape{cornercopoint}{
\inheritsavedanchors[from=rectangle] % this is nearly a rectangle
\inheritanchorborder[from=rectangle]
\inheritanchor[from=rectangle]{center}
\inheritanchor[from=rectangle]{north}
\inheritanchor[from=rectangle]{south}
\inheritanchor[from=rectangle]{west}
\inheritanchor[from=rectangle]{east}
\backgroundpath{% this is new
\southwest \pgf@xa=\pgf@x \pgf@ya=\pgf@y
\northeast \pgf@xb=\pgf@x \pgf@yb=\pgf@y

\pgfmathsetmacro{\pgf@shorten@left}{\pgfkeysvalueof{/tikz/shorten left}}
\pgfmathsetmacro{\pgf@shorten@right}{\pgfkeysvalueof{/tikz/shorten right}}

\pgfpathmoveto{\pgfpoint{0.5 * (\pgf@xa + \pgf@xb)}{\pgf@yb + 5pt}}
\pgfpathlineto{\pgfpoint{\pgf@xa - 8pt + \pgf@shorten@left}{\pgf@ya + 1.5 * \pgf@shorten@left}}
\pgfpathlineto{\pgfpoint{\pgf@xa - 8pt + \pgf@shorten@left}{\pgf@ya}}
\pgfpathlineto{\pgfpoint{\pgf@xb + 8pt - \pgf@shorten@right}{\pgf@ya}}
\pgfpathlineto{\pgfpoint{\pgf@xb + 8pt - \pgf@shorten@right}{\pgf@ya + 1.5 * \pgf@shorten@right}}
\pgfpathclose
}
}

\pgfdeclareshape{langpoint}{
\inheritsavedanchors[from=rectangle] % this is nearly a rectangle
\inheritanchorborder[from=rectangle]
\inheritanchor[from=rectangle]{center}
\inheritanchor[from=rectangle]{north}
\inheritanchor[from=rectangle]{south}
\inheritanchor[from=rectangle]{west}
\inheritanchor[from=rectangle]{east}
\backgroundpath{% this is new
\southwest \pgf@xa=\pgf@x \pgf@ya=\pgf@y
\northeast \pgf@xb=\pgf@x \pgf@yb=\pgf@y

\pgfmathsetmacro{\pgf@shorten@left}{\pgfkeysvalueof{/tikz/shorten left}}
\pgfmathsetmacro{\pgf@shorten@right}{\pgfkeysvalueof{/tikz/shorten right}}

\pgfpathmoveto{\pgfpoint{0.5 * (\pgf@xa + \pgf@xb)}{\pgf@ya - 2pt}}
\pgfpathlineto{\pgfpoint{\pgf@xa - 8pt}{\pgf@yb - 3 * \pgf@shorten@left + 5pt}} 
\pgfpathlineto{\pgfpoint{\pgf@xa - 8pt}{\pgf@yb -1pt}}
\pgfpathlineto{\pgfpoint{\pgf@xb + 8pt}{\pgf@yb -1pt}}
\pgfpathlineto{\pgfpoint{\pgf@xb + 8pt}{\pgf@yb - 3 * \pgf@shorten@left + 5pt}}%NEW POINT
\pgfpathclose
}
}

\pgfdeclareshape{langcopoint}{
\inheritsavedanchors[from=rectangle] % this is nearly a rectangle
\inheritanchorborder[from=rectangle]
\inheritanchor[from=rectangle]{center}
\inheritanchor[from=rectangle]{north}
\inheritanchor[from=rectangle]{south}
\inheritanchor[from=rectangle]{west}
\inheritanchor[from=rectangle]{east}
\backgroundpath{% this is new
\southwest \pgf@xa=\pgf@x \pgf@ya=\pgf@y
\northeast \pgf@xb=\pgf@x \pgf@yb=\pgf@y

\pgfmathsetmacro{\pgf@shorten@left}{\pgfkeysvalueof{/tikz/shorten left}}
\pgfmathsetmacro{\pgf@shorten@right}{\pgfkeysvalueof{/tikz/shorten right}}

\pgfpathmoveto{\pgfpoint{0.5 * (\pgf@xa + \pgf@xb)}{\pgf@yb +0pt}}
\pgfpathlineto{\pgfpoint{\pgf@xa - 8pt}{\pgf@ya + 3 * \pgf@shorten@left - 5pt}} 
\pgfpathlineto{\pgfpoint{\pgf@xa - 8pt}{\pgf@ya + 1pt}}
\pgfpathlineto{\pgfpoint{\pgf@xb + 8pt}{\pgf@ya + 1pt}}
\pgfpathlineto{\pgfpoint{\pgf@xb + 8pt}{\pgf@ya + 3 * \pgf@shorten@left - 5pt}}%NEW POINT
\pgfpathclose
}
}

\pgfdeclareshape{langrect}{
\inheritsavedanchors[from=rectangle] % this is nearly a rectangle
\inheritanchorborder[from=rectangle]
\inheritanchor[from=rectangle]{center}
\inheritanchor[from=rectangle]{north}
\inheritanchor[from=rectangle]{south}
\inheritanchor[from=rectangle]{west}
\inheritanchor[from=rectangle]{east}
\backgroundpath{% this is new
\southwest \pgf@xa=\pgf@x \pgf@ya=\pgf@y
\northeast \pgf@xb=\pgf@x \pgf@yb=\pgf@y

\pgfmathsetmacro{\pgf@shorten@left}{\pgfkeysvalueof{/tikz/shorten left}}
\pgfmathsetmacro{\pgf@shorten@right}{\pgfkeysvalueof{/tikz/shorten right}}

\pgfpathmoveto{\pgfpoint{\pgf@xa - 8pt}{\pgf@ya + 3 * \pgf@shorten@left - 5pt}} 
\pgfpathlineto{\pgfpoint{\pgf@xa - 8pt}{\pgf@ya + 1pt}}
\pgfpathlineto{\pgfpoint{\pgf@xb + 8pt}{\pgf@ya + 1pt}}
\pgfpathlineto{\pgfpoint{\pgf@xb + 8pt}{\pgf@ya + 3 * \pgf@shorten@left - 5pt}}%NEW POINT
\pgfpathclose
}
}
\makeatother

\pgfkeyssetvalue{/tikz/shorten left}{0pt}
\pgfkeyssetvalue{/tikz/shorten right}{0pt}

\tikzstyle{kpoint common}=[draw,fill=white,inner sep=1pt,minimum height=4mm]

\tikzstyle{langstate}=[shape=langcopoint,shorten left=5pt,kpoint common,font=\footnotesize]
\tikzstyle{langeffect}=[shape=langpoint,shorten left=5pt,kpoint common,font=\footnotesize]
\tikzstyle{langstatedash}=[shape=langcopoint,dashed, shorten left=5pt,kpoint common,font=\footnotesize]
\tikzstyle{langeffectdash}=[shape=langpoint,dashed, shorten left=5pt,kpoint common,font=\footnotesize]
\tikzstyle{langbox}=[shape=langrect,shorten left=5pt,kpoint common,font=\footnotesize] 
\tikzstyle{kpoint}=[shape=cornerpoint,shorten left=5pt,kpoint common]
\tikzstyle{kpoint adjoint}=[shape=cornercopoint,shorten left=5pt,kpoint common]

\tikzstyle{kpoint conjugate}=[shape=cornerpoint,shorten right=5pt,kpoint common]
\tikzstyle{kpoint transpose}=[shape=cornercopoint,shorten right=5pt,kpoint common]
\tikzstyle{kpoint symm}=[shape=cornerpoint,shorten left=5pt,shorten right=5pt,kpoint common]

\tikzstyle{black kpoint}=[shape=cornerpoint,shorten left=5pt,kpoint common,fill=black,font=\color{white}]
\tikzstyle{black kpoint adjoint}=[shape=cornercopoint,shorten left=5pt,kpoint common,fill=black,font=\color{white}]
\tikzstyle{black kpointadj}=[shape=cornercopoint,shorten left=5pt,kpoint common,fill=black,font=\color{white}]

\tikzstyle{black dkpoint}=[shape=cornerpoint,shorten left=5pt,kpoint common,fill=black, doubled,font=\color{white}]
\tikzstyle{black dkpoint adjoint}=[shape=cornercopoint,shorten left=5pt,kpoint common,fill=black, doubled,font=\color{white}]
\tikzstyle{black dkpointadj}=[shape=cornercopoint,shorten left=5pt,kpoint common,fill=black, doubled,font=\color{white}]

\tikzstyle{kpointdag}=[kpoint adjoint]
\tikzstyle{kpointadj}=[kpoint adjoint]
\tikzstyle{kpointconj}=[kpoint conjugate]
\tikzstyle{kpointtrans}=[kpoint transpose]

\tikzstyle{big kpoint}=[kpoint, minimum width=1.2 cm, minimum height=8mm, inner sep=4pt, text depth=3mm]

\tikzstyle{wide kpoint}=[kpoint, minimum width=1 cm, inner sep=2pt]%, text depth=-0.7 mm]
\tikzstyle{wide kpointdag}=[kpointdag, minimum width=1 cm, inner sep=2pt]%, text depth=0.7 mm]
\tikzstyle{wide kpointconj}=[kpointconj, minimum width=1 cm, inner sep=2pt]%, text depth=-0.7 mm]
\tikzstyle{wide kpointtrans}=[kpointtrans, minimum width=1 cm, inner sep=2pt]%, text depth=0.7 mm]

\tikzstyle{gray kpoint}=[kpoint,fill=gray!50!white]
\tikzstyle{gray kpointdag}=[kpointdag,fill=gray!50!white]
\tikzstyle{gray kpointadj}=[kpointadj,fill=gray!50!white]
\tikzstyle{gray kpointconj}=[kpointconj,fill=gray!50!white]
\tikzstyle{gray kpointtrans}=[kpointtrans,fill=gray!50!white]

\tikzstyle{gray dkpoint}=[kpoint,fill=gray!50!white,doubled]
\tikzstyle{gray dkpointdag}=[kpointdag,fill=gray!50!white,doubled]
\tikzstyle{gray dkpointadj}=[kpointadj,fill=gray!50!white,doubled]
\tikzstyle{gray dkpointconj}=[kpointconj,fill=gray!50!white,doubled]
\tikzstyle{gray dkpointtrans}=[kpointtrans,fill=gray!50!white,doubled]

\tikzstyle{white label}=[draw,fill=white,rectangle,inner sep=0.7 mm]
\tikzstyle{gray label}=[draw,fill=gray!50!white,rectangle,inner sep=0.7 mm]
\tikzstyle{black label}=[draw,fill=black,rectangle,inner sep=0.7 mm]

\tikzstyle{dkpoint}=[kpoint,doubled]
\tikzstyle{wide dkpoint}=[wide kpoint,doubled]
\tikzstyle{dkpointdag}=[kpoint adjoint,doubled]
\tikzstyle{wide dkpointdag}=[wide kpointdag,doubled]
\tikzstyle{dkcopoint}=[kpoint adjoint,doubled]
\tikzstyle{dkpointadj}=[kpoint adjoint,doubled]
\tikzstyle{dkpointconj}=[kpoint conjugate,doubled]
\tikzstyle{dkpointtrans}=[kpoint transpose,doubled]

\tikzstyle{kscalar}=[kpoint common, shape=EBox, inner xsep=-1pt, inner ysep=3pt,font=\small]
\tikzstyle{kscalarconj}=[kpoint common, shape=WBox, inner xsep=-1pt, inner ysep=3pt,font=\small]

 \tikzstyle{upground}=[circuit ee IEC,ground,rotate=90,scale=2.5]
 \tikzstyle{downground}=[circuit ee IEC,ground,rotate=-90,scale=2.5]
 \tikzstyle{bigground}=[regular polygon,regular polygon sides=3,draw=gray,scale=0.50,inner sep=-0.5pt,minimum width=10mm,fill=gray]
\tikzstyle{arrs}=[-latex,font=\small,auto]
\tikzstyle{arrow plain}=[arrs]
\tikzstyle{arrow dashed}=[dashed,arrs]
\tikzstyle{arrow bold}=[very thick,arrs]
\tikzstyle{arrow hide}=[draw=white!0,-]
\tikzstyle{arrow reverse}=[latex-]
\tikzstyle{cdnode}=[]

\definecolor{hexcolor0xa9a9a9}{rgb}{0.663,0.663,0.663}
\tikzstyle{GrayLine}=[dashed,draw=hexcolor0xa9a9a9]
\tikzstyle{gray}=[dashed,draw=hexcolor0xa9a9a9]

\theoremstyle{definition}
\newtheorem{theorem}{Theorem}[section]
\newtheorem*{theorem*}{Theorem}

\newtheorem{example*}[theorem]{Example*}
\newtheorem{examples*}[theorem]{Examples*}
\newtheorem{remark}[theorem]{Remark}
\newtheorem{remark*}[theorem]{Remark*}

\def\bR{\begin{color}{red}}
\def\bB{\begin{color}{blue}}
\def\bM{\begin{color}{magenta}}
\def\bC{\begin{color}{cyan}}
\def\bW{\begin{color}{white}}
\def\bBl{\begin{color}{black}}
\def\bG{\begin{color}{green}}
\def\bY{\begin{color}{yellow}}
\def\e{\end{color}\xspace}
\newcommand{\bit}{\begin{itemize}}
\newcommand{\eit}{\end{itemize}\par\noindent}
\newcommand{\ben}{\begin{enumerate}}
\newcommand{\een}{\end{enumerate}\par\noindent}
\newcommand{\beq}{\begin{equation}}
\newcommand{\eeq}{\end{equation}\par\noindent}
\newcommand{\beqa}{\begin{eqnarray*}}
\newcommand{\eeqa}{\end{eqnarray*}\par\noindent}
\newcommand{\beqn}{\begin{eqnarray}}
\newcommand{\eeqn}{\end{eqnarray}\par\noindent} 

\begin{document} 

\title{High schoolers excel at Oxford quantum course\\ using pictorial mathematics}

\author{Bob Coecke$^{1,9,10}$, Aleks Kissinger$^{2}$, Stefano Gogioso$^{2,3}$,
Selma D\"undar-Coecke$^{1, 4}$, Caterina Puca$^{1}$, Lia Yeh$^{1, 2}$,
Muhammad Hamza Waseem$^{1,5}$, Emmanuel M. Pothos$^{6}$, Sieglinde Pfaendler$^{7}$,
Vincent Wang-Mascianica$^{1}$, Thomas Cervoni$^{1}$, Ferdi Tomassini$^{1}$,
Vincent Anandraj$^{1}$, Peter Sigrist$^{1}$ and Ilyas Khan$^{1}$}

\affil{$^1$Quantinuum, 17 Beaumont Street, Oxford OX1 2NA, United Kingdom\vspace{1mm}}

\affil{$^2$Department of Computer Science, University of Oxford, United Kingdom\vspace{1mm}} 

\affil{$^3$Hashberg Ltd, 71-75 Shelton Street, WC2H 9JQ London, United Kingdom.\vspace{1mm}} 

\affil{$^4$Centre for Educational Neuroscience, University College London, United Kingdom.\vspace{1mm}} 

\affil{$^5$Department of Physics, University of Oxford, United Kingdom.\vspace{1mm}} 

\affil{$^6$Department of Psychology, City St. George's, University of London, United Kingdom.\vspace{1mm}} 

\affil{$^7$IBM Research GmbH, S\"aumerstrasse 4, CH–8803 R\"uschlikon, Switzerland.\vspace{1mm}} 

\affil{$^8$Moth Quantum, Somerset House, West Wing, WC2R 1LA London, United Kingdom.\vspace{1mm}} 

\affil{$^9$Perimeter Institute, 31 Caroline Street North, Waterloo, Ontario N2L 2Y5, Canada.\vspace{1mm}} 

\affil{$^{10}$Wolfson College, University of Oxford, United Kingdom.\vspace{2mm}} 

\affil{\ \ (affiliations are those at the time the work was done)}

\begin{abstract}
We are at the dawn of the second quantum revolution, where our ability to create and control individual quantum systems is poised to drive transformative advancements in basic science, computation, and everyday life. However, quantum theory has long been conceived as notoriously hard to learn, creating a significant barrier to workforce development, informed decision-making by stakeholders and policymakers, and broader public understanding. 

This paper is concerned with \textit{Quantum Picturalism}~\cite{ContPhys}, a novel visual mathematical language for quantum physics. Originally developed over two decades ago to explore the foundational structure of quantum theory~\cite{AC1}, this rigorous diagrammatic framework has since been adopted in both academia and industry as a powerful tool for quantum computing research and software development. Here, we demonstrate its potential as a transformative educational methodology.

We report the findings from a pilot study involving 54 UK high school students~\cite{dundar2023quantum}, randomly selected from a pool of 734 volunteers across the UK. Despite the absence of advanced mathematical prerequisites, these students demonstrated a strong conceptual grasp of key quantum principles and operations. On an assessment comprising university graduate-level exam questions, participants achieved an 82\% pass rate, with 48\% obtaining a distinction-level grade.   %On an assessment comprising university graduate-level exam questions, participants achieved a 72\% pass rate, with 37\% obtaining a distinction-level grade.  \bM After pre-filtering by using a diagnostic canary question, these figures improved to 82\% and 48\% respectively.\e
 
These results pave the way for making quantum more inclusive, lowering traditional cognitive and demographic barriers to quantum learning. This approach has the potential to broaden participation in the field and provide a promising new entry point for stakeholders, future experts, and the general public.
\end{abstract}

\vspace{-4mm}

\noindent
\emph{This paper is a contribution to a volume  in memory of Basil Hiley, who was one of the 1st fans of QPic, and likely the first person trying to teach it to their very young granddaughter, which happened around 2005.  Basil could see well beyond the current standards of physics, in a way very few can do.}

\section{Introduction}

Quantum theory is perhaps the greatest success story of 20th century physics, with our understanding of quantum effects playing a crucial role in everything from the atomic clocks used in GPS navigation to the transistors that make up every computer in the world. The emergence of quantum computation, wherein quantum systems themselves are used to store and process information, is expected to have a dramatic impact in the coming years in fields such as chemistry~\cite{cao2019quantum}, microbiology~\cite{pal2024future}, medicine~\cite{jeyaraman2024revolutionizing}, material science~\cite{bauer2020quantum}, communications~\cite{van2014quantum, orieux2016recent}, and artificial intelligence (AI)~\cite{dunjko2018machine}. Governments have already declared quantum technologies to be of the highest geopolitical importance~\cite{Geopolitics}, and quantum technologies have been predicted to have a huge economic impact in several major industries in the coming years \cite{Forbes, McKinsey}.

%The language in which quantum is traditionally formulated, following von Neumann \cite{vN}, is the language of complex Hilbert spaces, tensor products thereof, and linear maps between those (HilbS). This subject is usually taught in advanced undergraduate level mathematics and physics courses. Applications that lead to new quantum technologies are mostly taught at post-graduate level.

At the dawn of this transformative technological revolution, it is imperative to cultivate a skilled workforce, empower stakeholders in business and governance to make informed  and responsible decisions, and foster a society that is broadly quantum-ready across different age groups and different educational and socioeconomic backgrounds.

% We can contrast this with the current situation in AI technology, where society nor the stakeholders understand what is going on, as even the specialists and business leaders involved cannot even explain the workings of their systems.

Unfortunately, quantum theory is often regarded as only comprehensible for those with an exceptional intellect and who have had the opportunity to complete one or more advanced science degrees. The language in which quantum theory is traditionally formulated, following von Neumann~\cite{vN}, is that of complex Hilbert spaces and linear maps. Throughout this paper, we will refer to this traditional language as the Hilbert space formalism. The foundations of this language are usually taught in advanced undergraduate-level mathematics and physics courses, with advanced topics and applications, \textit{e.g.} in quantum technologies, often appearing only at postgraduate level.

Note that by `language' here we mean how quantum systems are described, how one reasons about them, and how calculations are done. Interestingly, von Neumann himself denounced the Hilbert space formalism merely three years after publishing his book on it~\cite{von2005john} and dedicated much time thereafter to finding an alternative language~\cite{Redei1}. 

This article is concerned with a new language for quantum theory and its applications, which we refer to as \emph{Quantum Picturalism} (QPic)~\cite{ContPhys}. It is the subject of two books written by some of the authors: \textit{Picturing Quantum Processes}~\cite{CKbook}, a textbook which has been used as the basis of a postgraduate course at Oxford University for about a decade; and \textit{Quantum in Pictures}~\cite{QiP}, a new accessible book introducing QPic to a broad audience, including young people and readers without any formal mathematical background.

Notably, \textit{Quantum in Pictures} has no mathematical prerequisites beyond what is already taught to 7-11 year olds in the UK, namely the ability to express angles in degrees and add them together~\cite{twinkle}. Still, the book covers advanced quantum topics, including some that have been discovered only in the past few decades, or even in just the past few years.

Initially, QPic was not intended to be an educational tool, but rather an abstract language for exploring the fundamental principles of quantum theory~\cite{AC1} and quantum computing \cite{CD1}. However, it has since found many applications in solving concrete problems in quantum technologies, such as optimising quantum computations to use less time and space~\cite{clifford-simp,KissingerTcount,cowtan2019phase,de2020fast,vandaele2024qubitcount,borgna2021hybrid,gogioso2023annealing,holker2023causal}, modelling and simulating quantum processes on a classical (super)computer~\cite{kissinger2022classical,wille2022basis,codsi2022classically,ufrecht2023cutting,cam2023speeding,KoziellPipe2024Simulation}, producing effective techniques to manage and correct errors in quantum computations~\cite{Gidney2019,litinski2022active,kissinger2022grok1,khesinGraphicalQuantumCliffordencoder2023,kissinger2024grok2,bombin2023unifying,townsend-teague2023floquetifying,wu2023zxcalculus,Huang2023,burton2024genons,boriskhesin2024equivalence,magdalenadelafuente2024xyz,rodatz2024floquetifying}, and designing interpretable classical and quantum AI systems~\cite{Nature, kartsaklis2021lambeq, qdisccirctheory, qdisccircexperiment}.

%Meanwhile, also in several other STEM areas there now are quantum picturalism alike languages being developed, which have the same potential, both for education and technology development \bR \cite{XXX}\e.\TODOb{Put refs.}

In this article, we provide evidence that QPic enables one to teach advanced quantum concepts at the high school level, and that moreover, the students can use that knowledge to solve sophisticated problems. We demonstrate that, after a short course, a group of 54 high school students, selected randomly from a pool of 734 volunteers from across the UK, could not only pass an exam based on past exam questions from an Oxford postgraduate-level course, but that many of them excelled. Of the students that participated in the study, 82\% passed the exam and 48\% obtained a distinction, \textit{i.e.},~a first-class mark of 70 or above.\footnote{These results were obtained after pre-filtering students by means of a ``canary question", which was designed to measure basic engagement of participants with the course material. That is, a canary question was a question which tested participants' overall understanding; correct answers to the canary question (it had two parts)  was a prerequisite for passing the assessment.%These results were obtained after pre-filtering students by means of a canary question, which ensured a fundamental understanding of core quantum concepts and operations. 
}   
These results are especially promising given that this was a completely voluntary course undertaken by UK high school students, many of whom were simultaneously preparing for their A-level examinations.~\cite{Alevel}. This supports our argument that QPic is a powerful new tool to make quantum education more inclusive than imagined, in a way that goes hand-in-hand with cutting-edge quantum technology development.

\section{Quantum Picturalism}

Quantum Picturalism (QPic) employs a new and highly intuitive form of mathematics known as \textit{diagrammatic reasoning}, departing from conventional symbolic methods. This mathematical framework was formalised in the 1990s and early 2000s through the theory of monoidal categories \cite{Benabou, KellyLaplaza, CarboniWalters, Lack}. However, the particular types of diagrams used trace back two decades earlier to the graphical tensor notation introduced by Penrose \cite{Penrose}. QPic shows that this pictorial notation can be extended, capturing all of the essential quantum notions like entanglement, mixed states and processes, and observables and their complementarity \cite{Kindergarten, SelingerCPM, CPaqPav, CPer, CD2, CDKZ}.

QPic originated as an attempt to provide a high-level language for quantum computation, aimed at helping quantum computing researchers better understand and design algorithms. The basic idea proceeds in analogy with the case of classical software. Whereas the very first classical computers were programmed using very rudimentary machine code, which got translated directly into streams of 1s and 0s, modern sophisticated software would not be possible without \textit{high-level} programming languages like C++ and Python. These languages allow programmers to describe complex behaviours in ways that more closely resemble how they think about problems, rather than focusing on the low-level details of the machine.

The analogy goes as follows. While the low-level language of classical computing consists of binary numbers, \textit{i.e.} matrices of 0's and 1's, the `low-level language' of quantum theory comprises matrices of complex numbers. QPic gives us a high-level language for describing quantum theory using pictures, \textit{e.g.}~from \cite{CKbook}:
\begin{center}
$\footnotesize\frac{1}{4}
\left(
\begin{array}{rrrrrrrr}
\!-\!1\!+\!i & 1\!+\!i & 1\!+\!i & \!-\!1\!+\!i & 1\!+\!i & 1\!-\!i & 1\!-\!i & 1\!+\!i \\
1\!+\!i & 1\!-\!i & 1\!-\!i & 1\!+\!i & \!-\!1\!+\!i & 1\!+\!i & 1\!+\!i & \!-\!1\!+\!i \\
1\!+\!i & 1\!-\!i & 1\!-\!i & 1\!+\!i & 1\!-\!i & \!-\!1\!-\!i & \!-\!1\!-\!i & 1\!-\!i \\
1\!-\!i & \!-\!1\!-\!i & \!-\!1\!-\!i & 1\!-\!i & 1\!+\!i & 1\!-\!i & 1\!-\!i & 1\!+\!i \\
1\!+\!i & 1\!-\!i & 1\!-\!i & 1\!+\!i & 1\!-\!i & \!-\!1\!-\!i & \!-\!1\!-\!i & 1\!-\!i \\
1\!-\!i & \!-\!1\!-\!i & \!-\!1\!-\!i & 1\!-\!i & 1\!+\!i & 1\!-\!i & 1\!-\!i & 1\!+\!i \\
\!-\!1\!+\!i & 1\!+\!i & 1\!+\!i & \!-\!1\!+\!i & 1\!+\!i & 1\!-\!i & 1\!-\!i & 1\!+\!i \\
1\!+\!i & 1\!-\!i & 1\!-\!i & 1\!+\!i & \!-\!1\!+\!i & 1\!+\!i & 1\!+\!i & \!-\!1\!+\!i \end{array}
\right)$\qquad\qquad $\Longrightarrow$\qquad \qquad\begin{tikzpicture}
	\begin{pgfonlayer}{nodelayer}
		\node [style=none] (0) at (-2, -2) {};
		\node [style=none] (1) at (0, -2) {};
		\node [style=none] (2) at (2, -2) {};
		\node [style=gray phase dot] (3) at (0, 0) {\scriptsize$\pi\over 2$};
		\node [style=gray phase dot] (4) at (1, -1) {\scriptsize$\pi\over 2$};
		\node [style=gray phase dot] (5) at (-1, 1) {\scriptsize$\pi\over 2$};
		\node [style=none] (6) at (2, 2) {};
		\node [style=none] (7) at (-2, 2) {};
		\node [style=none] (8) at (0, 2) {};
	\end{pgfonlayer}
	\begin{pgfonlayer}{edgelayer}
		\draw [style=swap, bend right=45, looseness=1.00] (7.center) to (5);
		\draw [style=swap, bend right=45, looseness=1.00] (5) to (8.center);
		\draw [style=swap, bend left=45, looseness=1.00] (1.center) to (4);
		\draw [style=swap, bend left=45, looseness=1.00] (4) to (2.center);
		\draw [style=swap, in=90, out=-150, looseness=0.75] (3) to (0.center);
		\draw [style=swap, in=-90, out=30, looseness=0.75] (3) to (6.center);
	\end{pgfonlayer}
\end{tikzpicture}
\end{center}
The matrix on the left and the picture on the right describe the same mathematical object. However, even without knowing the details of quantum theory, one can immediately recognise certain features from the picture, \textit{e.g.}~which parts are connected to others---that are completely obscured by the matrix.

%As also mentioned earlier, meanwhile QPict is widespread in most of the large quantum technology companies \cite{cowtan2019phase, Gidney2019, litinski2022active}, with also high-profile academics \bR like the inventor of modern quantum computing Peter Shor using it\e \cite{khesinGraphicalQuantumCliffordencoder2023}. Some example applications, IOOA, include making protocols involving entanglement easier to design and analyse \cite{LE1, AC1, Kindergarten}, designing and manipulating quantum circuits \cite{CD2} and other quantum computational models such as MBQC \cite{CD2, DP2}, error-correction \cite{Gidney2019, huang2023graphical, khesinGraphicalQuantumCliffordencoder2023}, all forms of photonics quantum computing \cite{defeliceQuantumLinearOptics2022, litinski2022active}, %There are Below in Sec.~\ref{Sec:ex} we discuss tasks that simply cannot be solved without QPict.

We now briefly give a taste of QPic. %\TODOb{This bit can be formatted more in the style of the book.}
The basic building blocks from which we built quantum states and processes are called \emph{spiders}. These spiders are of two colours and depicted as follows:
\[
\epsfig{figure=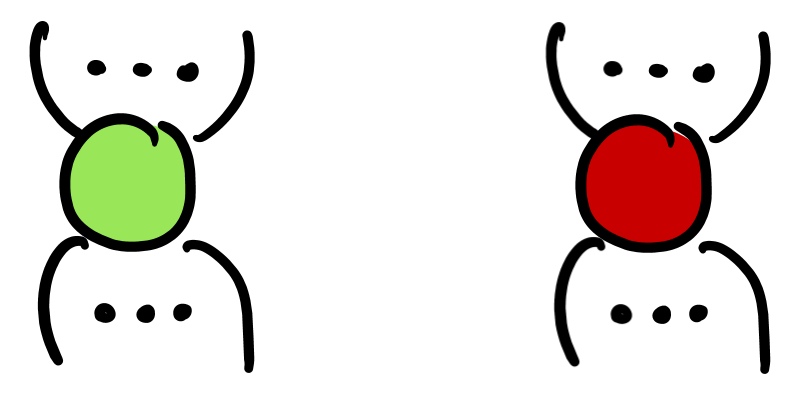,width=120pt}
\]
These can then be plugged together by \emph{connecting legs}:
\[
\epsfig{figure=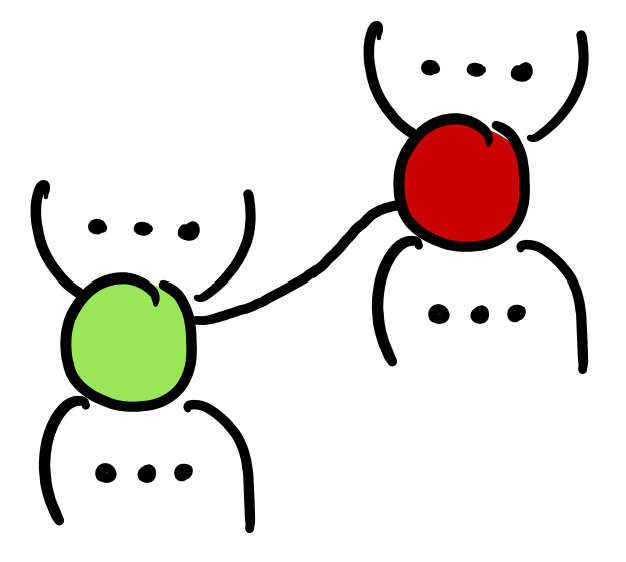,width=95pt}
\]
Connected spiders are subject to the following rules. If they are of the same colour, they \emph{fuse}:
\[
\epsfig{figure=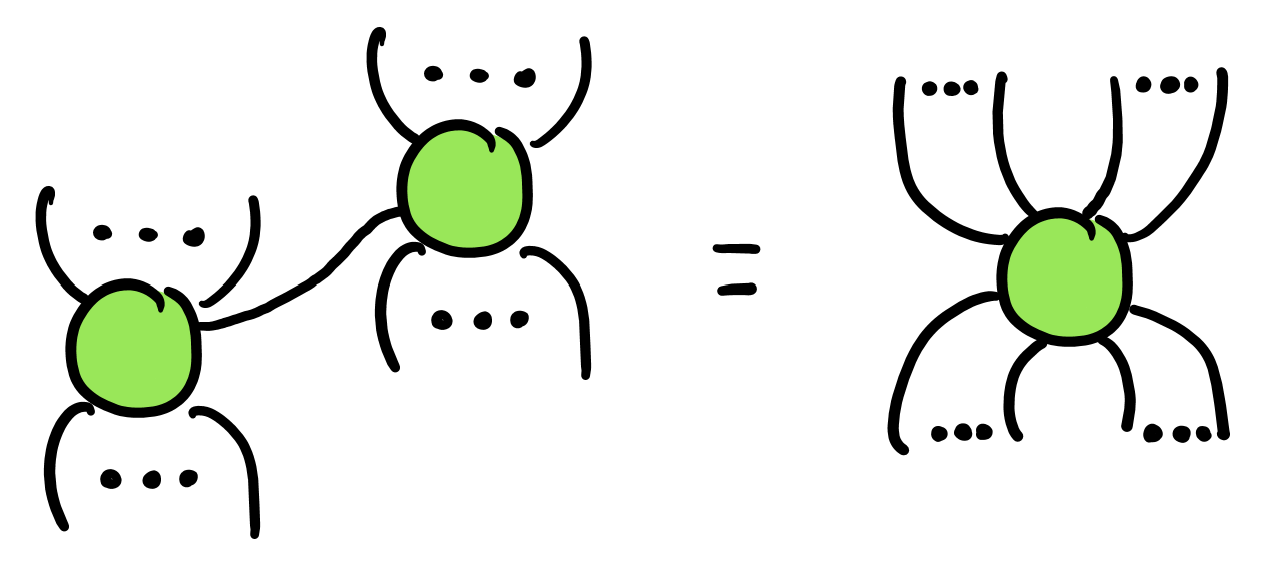,width=190pt}
\]
If they are of different colour and share two legs, we \emph{chop} those two legs:
\[
\epsfig{figure=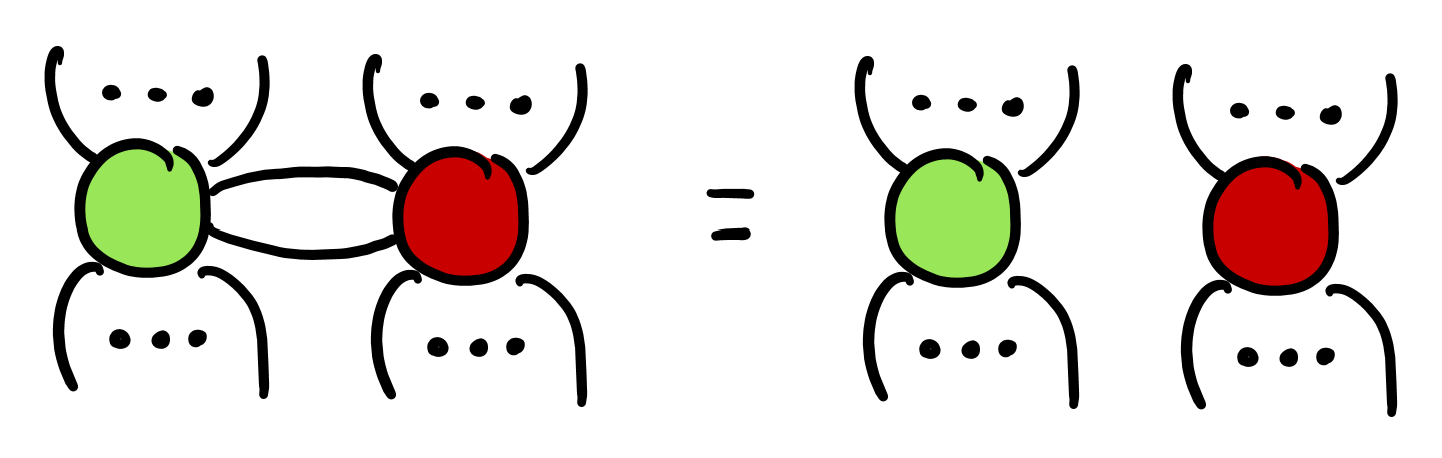,width=200pt}
\]

Now consider the following \emph{quantum circuit}, in which the vertical lines are \emph{qubits}---so there are three of those---and the horizontal lines with spiders at each end are \emph{CNOT-gates}---hence there are three of those:
\[
\epsfig{figure=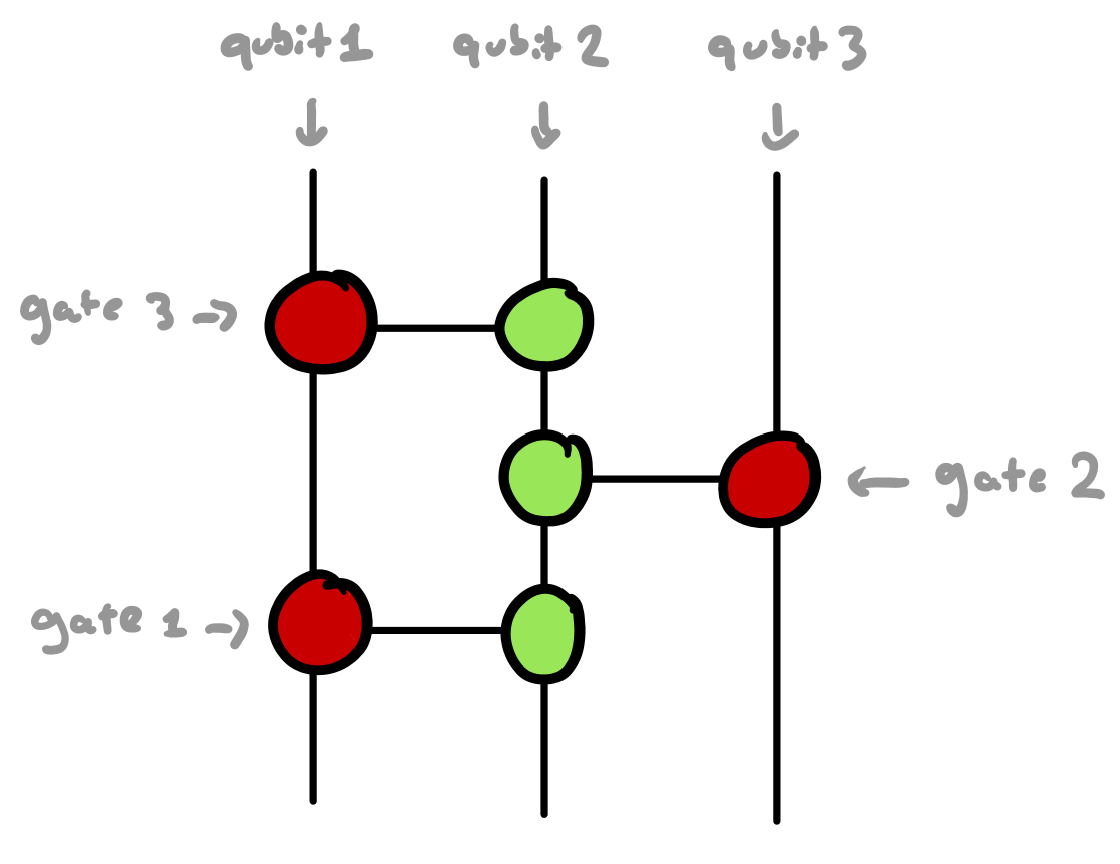,width=160pt}
\]
We immediately see that there are connected spiders of the same colour, so we can fuse them:
\[
\epsfig{figure=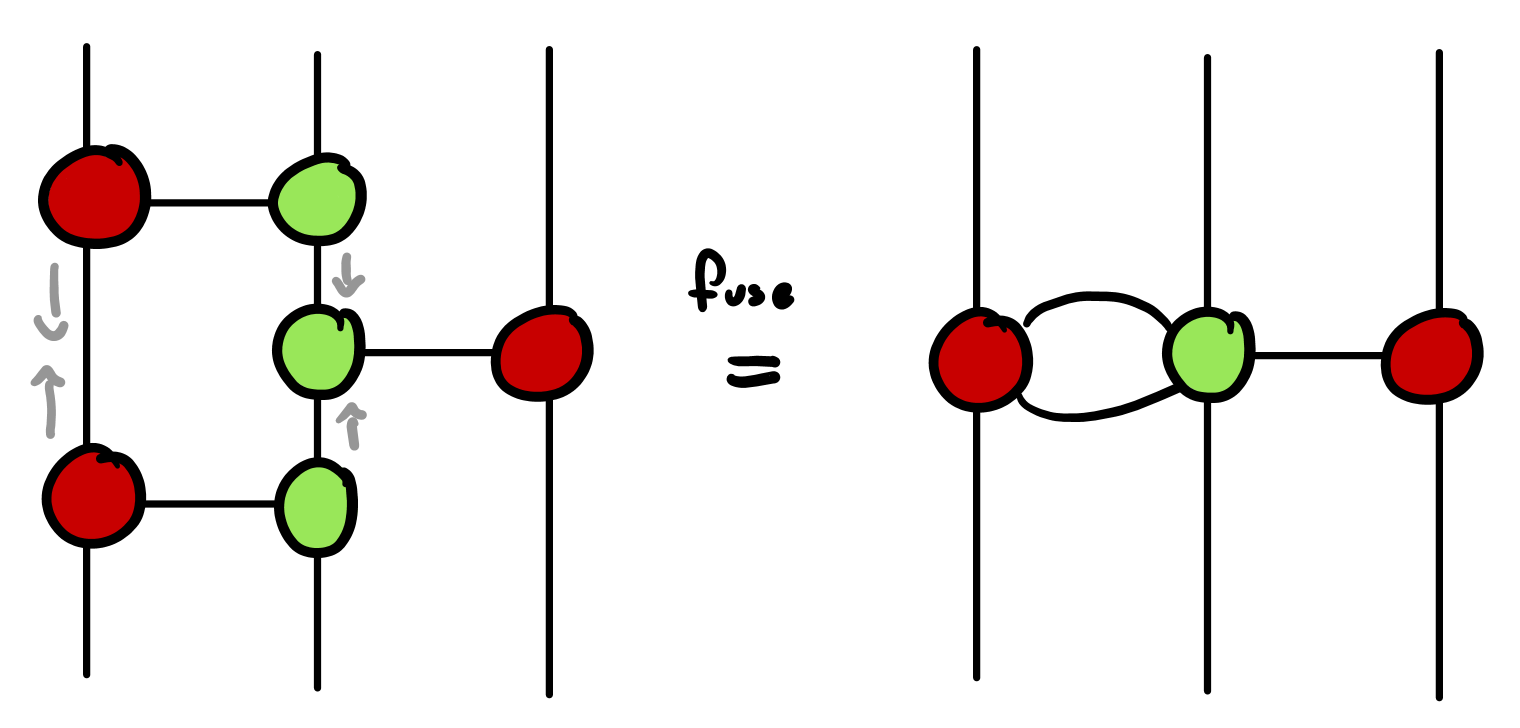,width=235pt}
\]
Next we see that there are two shared legs between two spiders of different colours, so they vanish:
\[
\epsfig{figure=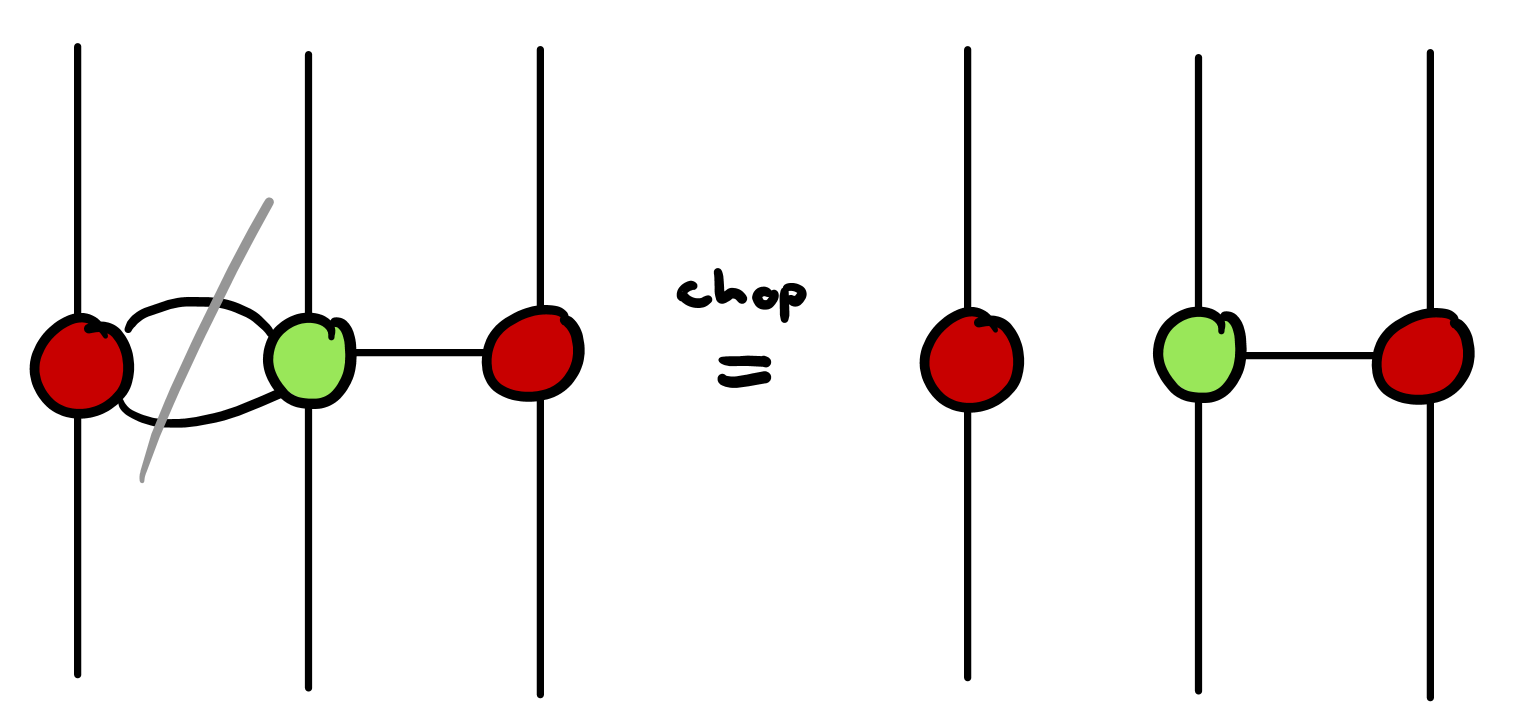,width=240pt}
\]
Spiders with two legs are just wires, therefore:
\[
\epsfig{figure=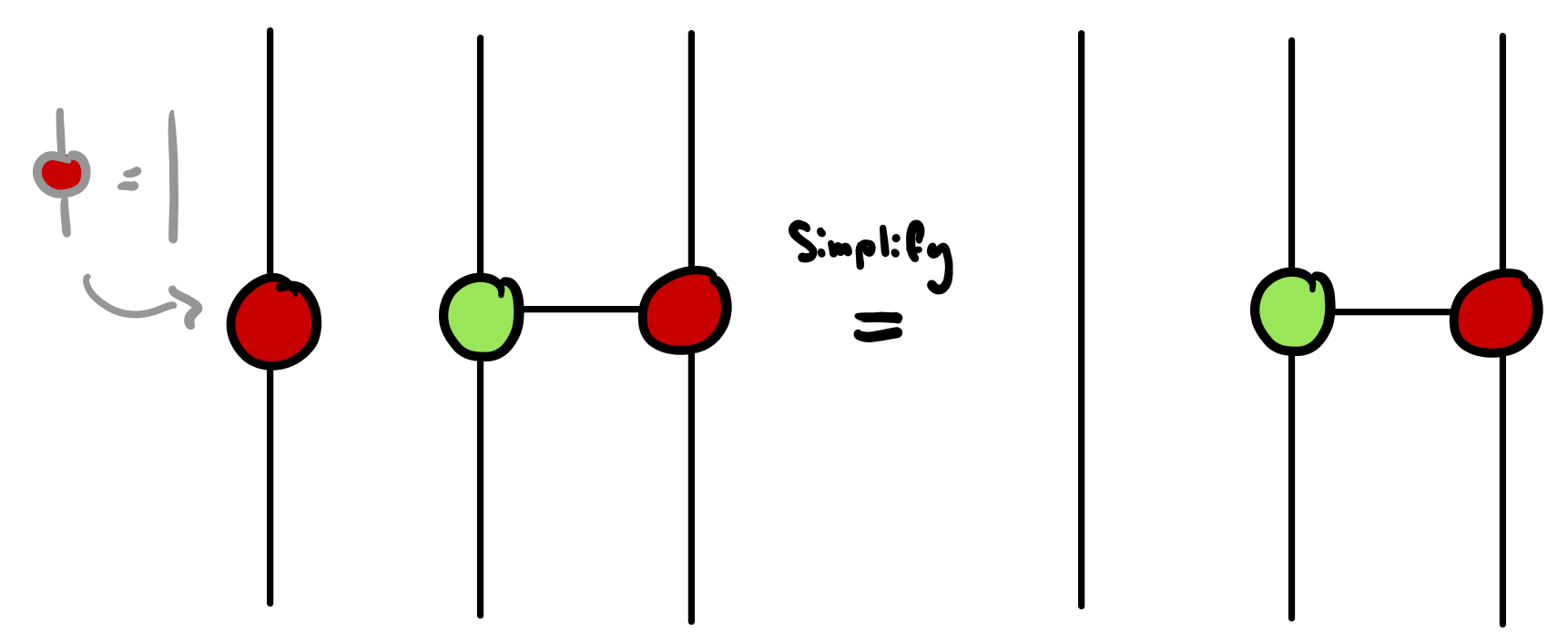,width=285pt} \qquad\ \
\]
Hence we reduced this quantum circuit from having three CNOT-gates to having only one CNOT-gate. This is of course a very simple example, but the methods based on simple diagram transformations like these have been used in state-of-the-art quantum circuit optimisation procedures \cite{clifford-simp, cowtan2019phase, de2020fast}. This is very important, as it allows large quantum programs to be turned into much smaller ones that can be executed faster and with few opportunities for errors. Other equally simple methods underpin many other important tools for quantum technologies.

How broadly applicable is QPic? For over 10 years, it has been the subject of a popular post-graduate course at Oxford University, containing all essentials expected from a quantum computing course \cite{CKbook}. In addition to five courses at Oxford University teaching QPic, it has also been taught in courses at the University of Edinburgh and the University of Cambridge in the UK, Indiana University Bloomington in the US, the University of Amsterdam in the Netherlands, Kwame Nkrumah University of Science and Technology in Ghana, and more courses listed in the extended version of this paper~\cite{dundarcoecke2025makingquantumworldaccessible}. QPic methods have been applied in roughly 300 research papers including publications by 10 quantum computing companies, across 18 areas: quantum error correction, optimisation, classical simulation, variational algorithms, verification, measurement-based quantum computation, natural language processing, photonic circuits, condensed matter, chemistry, higher-dimensional quantum computation, continuous-variable quantum computation, cryptography, entanglement, stabiliser theory, complexity theory, quantum computer architecture, and quantum software. The utility of QPic in research and teaching stems from the following results:
\bit
\item \emph{Universality} \cite{ContPhys}. All states and processes of the Hilbert space formalism can be represented within QPic.
\item \emph{Completeness} \cite{hadzihasanovic2018two, vilmart2018near}. Any equation between states and processes that can be derived within the Hilbert space formalism can also be derived within QPic. 
\eit

The above discussion shows that QPic diagrams are somewhat different in nature from other familiar diagrammatic tools, such as Feynman diagrams. Whereas the latter serve as a visual aid to guide rigorous calculations that must still be carried out using the `standard' mathematics of quantum theory, QPic is already a fully rigorous language in which calculations can be performed directly.

Importantly, in many cases, the representation and the derivations in QPic turn out to be much simpler than those in the Hilbert space formalism. How much simpler? This brings us to our main result of this paper.

\section{The course}

We describe the course materials and the content of the exam in this section. 
%Details about the experimental methodology are provided in Section~\ref{methodsect}. 
Additional background, context, methodology and more details of the study and the results reported in this paper can be found in the pre-experiment paper~\cite{dundar2023quantum} and the extended and more technical version of this paper \cite{dundarcoecke2025makingquantumworldaccessible}.  

\subsection{Course materials}

The course materials consisted of a course textbook, which was a slightly abridged version of the book \textit{Quantum in Pictures}, covering the following topics:
\bit%\COMMb{I went through the book for this list.} 
%\item general states, processes, their sequential composition (cf.~matrix multiplication) and parallel composition (cf.~tensor product), quantum and classical bits, unitary processes, inner-product, Bell states/effects, observables, non-determinism, measurement disturbance, probabilities, complementarity, causality and no-signalling, quantum teleportation, entanglement swapping, no-cloning and no-broadcasting, quantum key distribution, quantum circuits and circuit optimisation, measurement-based quantum computing, mixed states, completely positive trace-preserving (CPTP) maps, decoherence, quantum non-locality and the Greenberger–Horne–Zeilinger (GHZ) argument.
\item general states, processes, their sequential composition (cf.~matrix multiplication) and parallel composition (cf.~tensor product), process theories and Choi-Jamiołkowski isomorphism, spacetime diagrams, quantum and classical bits, quantum gates (CNOT, H, Z phase, X phase), quantum circuits, unitary processes, inner product, Bell states/effects, observables, non-determinism, measurement disturbance, complementarity, causality and no-signalling, quantum teleportation, entanglement swapping, reversible computation, entanglement, uncertainty principle, gate-based quantum computing, quantum circuit optimisation, measurement-based quantum computing, multi-party quantum communication protocols, mixed state quantum information

\eit
%\TODOa{this list seems a bit random. check this with what we actually covered}
In \textit{Quantum in Pictures}, many of these concepts were given alternative (and somewhat less obscure) names that more directly resemble their respective pictorial incarnations and, similarly, the subjects were given names that more directly relate to daily reality. A translation is provided at the end of \textit{Quantum in Pictures}.

\begin{remark}
Two notable omissions here are the Schr\"odinger equation and any detailed discussion of quantum algorithms. For the former, we follow the conventions of a typical quantum information/computation course and break time evolution down into discrete time steps, thus not requiring manipulation of differential equations. For the latter, producing small, digestible examples of problems that can be solved by a quantum algorithm suitable for education at this level is a topic of future work. We hope that this will build on the progress that has already been made in the analysis of quantum algorithms themselves using QPic~\cite{gogioso2017fully}.
\end{remark}

% \begin{remark}
% One topic that was not included and relevant for quantum technology are quantum algorithms. The reason is that the statement of the problem already requires more advanced mathematics or other scientific knowledge like chemistry or materials. However, the quantum algorithms themselves can be easily stated within QPict \cite{CKbook, gogioso2017fully}.
% \end{remark}

% \begin{remark}
% Also Schr\"odinger equation is not covered explicitly, as is the case in most theoretical quantum computation publications. Instead, time steps are discretised and Schr\"odinger equation is replaced by the unitarity restriction on gates. Also Schr\"odinger equation can be captured in QPict \cite{shaikh2022sum}.
% \end{remark}

%\bR YES/NO: Some subjects not covered include quantum algorithms, as the statement of the problem already requires more advanced mathematics or physics than we wanted to assume, and Schr\"odinger equation, as it typically doesn't occur in most theoretical quantum foundations and quantum computation discussions. Instead, time steps are discretised and Schr\"odinger equation is replaced by the unitarity restriction on pure processes. However, both subjects have been captured within QPict \bR\cite{CKbook, gogioso2017fully, X, X}\e.\e\TODOb{Do we want to put this here? Or a Q\&A at the end.}

In addition to the textbook, we also filmed blackboard lectures covering all the relevant material. Additional materials also included exercises with filmed solutions. All of these soon will be made publicly available, after post-processing. Here is a pre-production example of a lecture:%\COMMb{These will soon be public.}

{\footnotesize\begin{center}
\href{https://www.youtube.com/watch?v=uVdWzQwAO48}{https://www.youtube.com/watch?v=uVdWzQwAO48}
\end{center}}
\noindent
of an exercise:
{\footnotesize\begin{center}
\href{https://forms.office.com/pages/responsepage.aspx?id=G96VzPWXk0-0uv5ouFLPkQY1yz0RM0lIrQkpLDzPs4ZUOFA3N0k3Rlg3MVhQTzNCM0NUSlI0OEFJMy4u}{https://forms.office.com/pages/responsepage.aspx?id=G96VzPWXk0-0uv5ouFLPkQY1yz0RM0lIrQkpLDzPs4ZUOFA3N0k3Rlg3MVhQTzNCM0NUSlI0OEFJMy4u}
\end{center}}
\noindent
and the solutions:
{\footnotesize\begin{center}
\href{https://drive.google.com/drive/folders/18T54TfSCNuTOxJ6OHpGo6omGAE-17JEv}{https://drive.google.com/drive/folders/18T54TfSCNuTOxJ6OHpGo6omGAE-17JEv}
\end{center}}

%\subsection{Participant recruitment}

\subsection{Course structure}

The structure of the course was based roughly on that of an Oxford University postgraduate course. This consisted of an eight-week taught portion, with lectures, tutorials, and weekly exercises, followed by a take-home exam (which in Oxford post-graduate courses is sometimes called a \textit{mini-project}).

Students were asked to watch one video lecture per week in preparation for the weekly live tutorials. During the hour-long tutorial sessions, conducted in groups of up to 15 students, a tutor facilitated discussions, allowing participants to ask questions about the material and work through exercises under the tutor's guidance. Twelve additional homework exercises, with the filmed solutions, were also released during these 8 weeks, and the students were given about a week to try to do them, before solutions were released.

\subsection{The exam}\label{Sec:ex}

The structure of the exam closely mirrored that of an Oxford University postgraduate examination. It was double-marked, and in cases where the marks deviated by more than nine points, a third marker was involved. The exam followed a take-home assignment format, allowing participants three weeks to complete their assignments. With the exception of  a two-part canary question, all questions were adapted from previous Oxford postgraduate exams, modified only slightly to align with the terminology and notation taught during the eight-week course. The exam consisted of three multi-part questions, each progressively increasing in difficulty. %Here is the exam: \bR [EXAM LINK (PRIVATE?)]\e.

For example, participants were given the following table of rules to use to calculate their solutions:
\[
%\mbox{\bR...fusion, copy, leg-chop, colour change, square-pop all with names...\e}
\epsfig{figure=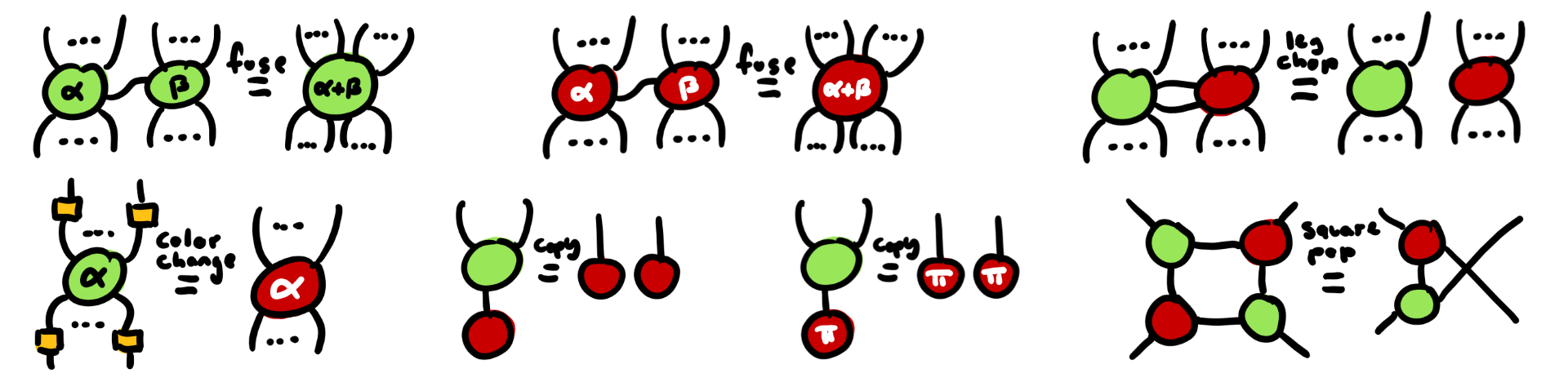,width=460pt}
\]

Below is an outline of the exam questions, an indication of their expected difficulty, and the QPic rules that needed to be used:
\bit
 \item
 The first question concerned quantum circuit simplification. Question 1a required the application of a four-qubit circuit to a state. It can be solved fairly straightforwardly both in QPic (using fusion and copy) and in Hilbert space formalism (applying gates one-by-one). Question 1b required showing that a three-qubit circuit reduces to the identity. In QPic it requires some more rules (leg-chop and colour-change), and in Hilbert space formalism multiplying 8x8 matrices for verification which is a lot more tedious, and easily leads to mistakes. Question 1c asks what a three-qubit circuit simplifies to. In QPic it requires one more rule (square-pop), but in Hilbert space formalism this question would again require a sizeable matrix calculation.
\item
The second question concerned multi-party protocols, more specifically, how a number of parties can together prepare a joint state using local operations and classical communication. Questions 2a and 2b are much more intuitive to solve with QPic than in the Hilbert space formalism. Question 2c is very hard in the Hilbert space formalism, as it requires an impossibility argument.
\item
The third question concerns computing classical correlations when measuring quantum states. The first two parts of Question 3, 3a.1 and 3a.2 are extremely straightforward in QPic, and essentially required them to reproduce two definitions given in the course. We deliberately included this two-part `canary' question to identify papers where students made very little effort to read the material and answer the questions.
% Concretely, the question simply asked for two spiders to either be fused, and to be leg-chopped respectively:
% \[
% \epsfig{figure=figs/Figure10,width=210pt}
% \]
Questions 3a.3 and 3a.4, on the other hand, were more difficult. To solve them properly, they require mixed states in the Hilbert space formalism, as they involve ``tracing out'' quantum systems, while in QPic they only require the much easier \emph{doubling} of diagrams, which then allows one to just stick to the same rules. Question 3b is difficult even in QPic, so gives us an indication of which students gained an exceptionally strong grasp of the material. We refer to this as the `sentinel' question.
\eit

\section{Results}  

Discarding those exams from students who failed to solve the two-part canary question, we found that 82\% of the students passed, of which 48\% had a distinction.\footnote{Readers interested in a comprehensive analysis of the experimental results are referred to the longer, more detailed paper~\cite{dundarcoecke2025makingquantumworldaccessible}.} 

There were several limitations in the course format during the pilot study, which, if addressed, could lead to even better student performance and experience. The first was that all course components were delivered online, with lectures and exercise solutions provided as pre-recorded materials. Due to institutional restrictions on experiments involving minors, participants had no opportunity to interact with each other outside tutorial sessions, such as through online platforms. An in-person course or an online format with enhanced peer interaction could foster collaboration, enabling participants to support one another and engage more naturally with tutors. Finally, the course timing posed a challenge, as it coincided with a busy exam season followed by family holiday period in the UK. A different schedule may have allowed participants to have fewer competing commitments.

\begin{figure}[H]
    \centering
    
    % Pie chart
    \begin{subfigure}[b]{0.45\textwidth}
        \centering
        \includegraphics[scale=0.4, trim = 10mm 10mm 10mm 10mm]{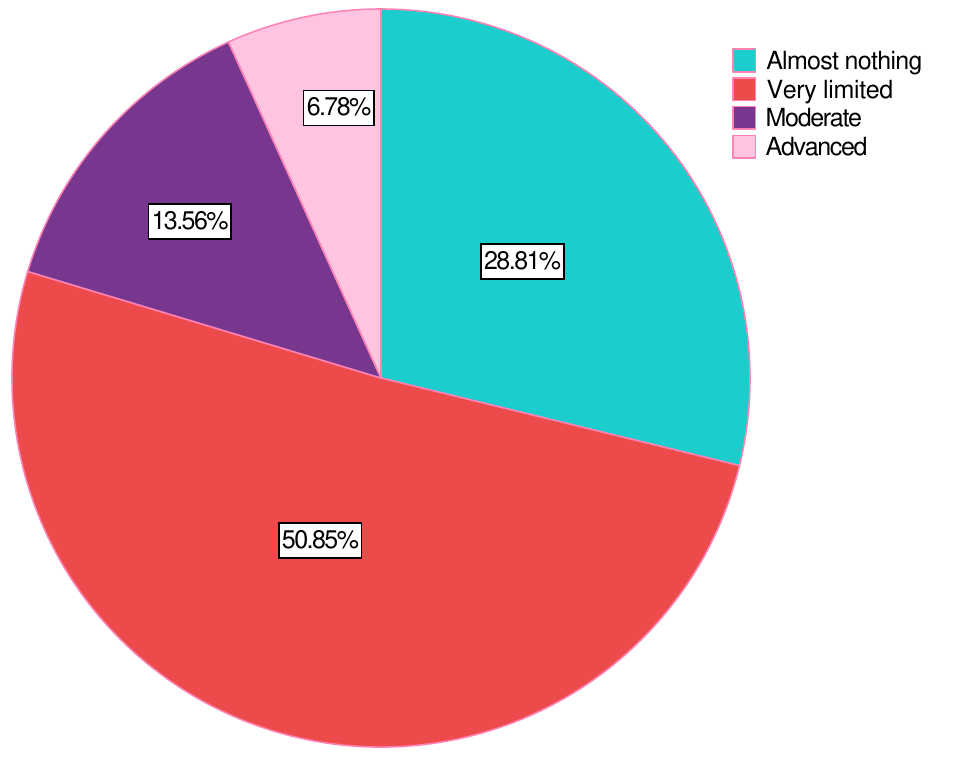}
        \vspace{0.8em}
        \centering
        \caption*{\textbf{(a) Prior knowledge of quantum}}
        \label{fig:pie}
    \end{subfigure}
    \hfill
    % First table
    \begin{subfigure}[b]{0.45\textwidth}
        \centering
        \raisebox{1.5em}{%
        \parbox{\linewidth}{%
        \centering
        \caption*{\textbf{(b) Hours spent to study per week.}}
        \vspace{0.3em}
        \begin{tabular}{ll}
            \toprule
            \emph{Hours/week} & \emph{Percent} \\
            \midrule
            1-2 & 64.4 \\
3-4 & 30.5 \\
4-5 & 1.7 \\
\emph{Total} & \emph{96.6} \\
\bottomrule
            
        \end{tabular}
        %\caption{Hours spent to study per week}
        }%
    }
        \label{table:hours}
    \end{subfigure}
    \hfill
    \vspace{1em}
    % Second table
    \begin{subfigure}[t]{0.9\textwidth}
        \centering
        \caption*{\textbf{(c) Other commitments over the course period.}}
        \vspace{0.3em}
        \begin{tabular}{cccccc}
            \toprule
            & \emph{Exam preparation}
& \emph{Part-time job}
& \emph{Internship}
& \emph{Sport activity}
& \emph{Camps/leisure} \\
\midrule
\% & 100 & 20.3 & 11.9 & 10.2 & 57.6 \\
\bottomrule
        \end{tabular}
        \label{table:commits}
    \end{subfigure}

    \caption{\label{fig:priorknowledge}Panel combining (a) A pie chart describing students' prior knowledge of quantum before attending training. The students were asked: “How much did you know about quantum theory before participating in this program?” 4 in 5 students responded either ‘Very limited’ or ‘Almost nothing’. (b) A table reporting study hours per week, and (c) A table reporting other commitments over the course period.}
    \label{fig:panel}
\end{figure}

 Students completed a questionnaire at the end of the course.  
Figure~\ref{fig:priorknowledge} presents a composite view of some of the data collected. Panel (a) shows that almost 80\% students reported limited to minimal prior knowledge of quantum. Panel (b) provides a breakdown of the number of hours participants reported studying per week, with almost two-thirds reporting study hours within the bounds of course expectations. Panel (c) displays all the other responsibilities the students were balancing during the course. As shown in the panel, the sample reflects a broad cross-section of students, indicating that it included varying levels of academic background, effort and engagement. Among the 54 students, more than half (55.6\%) attended state schools, which are funded by the government and usually do not require entrance exams.  25.9\% attended grammar schools, 5.6\% attended state boarding schools, 3.7\% attended private boarding schools, 3.7\% attended private independent schools, and 5.5\% (3 participants) opted not to respond. Further tables and detailed discussion of this data can be found in the extended version of this paper~\cite{dundarcoecke2025makingquantumworldaccessible}.
\vspace{1.5em}
%\begin{comment}

%\end{comment}

%\begin{figure}[h!]
 % \centering
  %\includegraphics[width=90mm]{priorknowledge.pdf}
  %\caption{\label{fig:priorknowledge} The students were asked: ``How much did you know about quantum theory before participating in this program?'' 4 in 5 students responded either `Very limited' or `Almost nothing'.}
%\end{figure}

The students' achievement was assessed through an end-of-course exam, which is described in Section~\ref{Sec:ex}. The exams were marked by University of Oxford faculty members according to standard procedures of the university.

%The mean exam scores by school type are presented in Figure~\ref{fig:meanperformance}. This is with 94.9\% of the students spending no more than 4 hours studying for the course per week, with 64.4\% spending only 1-2 hours per week. 

Importantly, the students also enjoyed the course: 86.4\% of students agreed with the statement, ``The course motivated me to pursue a STEM career,'' with 54.2\% strongly agreeing.

%\begin{figure}[h!]
  %\centering
  %\includegraphics[width=90mm]{meanperformance.pdf}
  %\caption{\label{fig:meanperformance}  \textcolor{pink}{Mean performance on the course final exam, demonstrating students performed well across all school types.}}
%\end{figure}

% The fact that only \bR 12\e students passed the sentinel question confirms that the students were fairly randomly chosen. Overall the scores were very polarised. There were clearly three groups: 1) the lowest performers, who could not complete the canary question; 2) those around 40-60 mark, that clearly had invested some effort in studying the material, and a majority passed (\bR40\%\e); 3) those above 70 mark, and most of them could clearly handle even more challenging questions than those in the exam.

% Furthermore, we decided early on to avoid potentially pressuring students to do well and award all students with the same certificate, regardless of their performance. Introducing more influential consequences for good and bad performance could also push scores up.

% The students didn't have any direct consequences if they performed badly, as we gave a certificate to everyone that took the course. (5) The period of teaching included A-level exams, regular school, and family summer holidays. (6) The tutors found it particularly hard to teach this generation, who grew up with smartphones in a Covid-era. Hardly any student used camera, and questions were typed.

% \bR ...something on the fact that younger people did better...\e
%\bR ... MORE CAN BE SAID HERE ... MAYBE SOME PICTURES WITH DATA POINTS? ...\e

\section{Impact}

The students' mastery of the key quantum theory concepts we tested exceeded our expectations, demonstrating promise that it is possible to teach a subject which has historically had a high barrier to entry. The significance of these results warrants further studies, particularly those featuring in-person teaching, which may further enhance comprehension and engagement. Preparations for these forthcoming studies are underway in the UK and internationally. International interest has been substantial since the very beginning, with many applications received from outside the UK to participate in this first trial, which were excluded in order to comply with Oxford University's ethical approval requirements. At the time of writing, some follow-up studies are being planned with US funding agencies in collaboration with the Colorado high school system, while others are being discussed in developing countries, such as South Africa, Ghana, Morocco, and Pakistan. These further studies are expected to provide deeper insights into the demographic and age groups that may benefit most, from utilising QPic to teach quantum concepts more accurately and more accessibly.

On a broader scale, QPic is being considered for systematic integration in educational curricula, both at the high school and university level, with plans with the Colorado high school system to examine the quantum concepts learnable by even younger age groups. In Greece, the ``KYMA (Quantum Computing for Students)" \cite{kyma} initiative was launched in 2021 with the goal of bringing high school students in contact with the world of Quantum Information. Their 2024-2025 program -- now engaging schools nationwide in Greece --  has incorporated elements of QPic and at the time of writing, plans are underway to further expand its integration in future realizations. The KYMA initiative follows up on Hellenic Governmental Authorities’ official approval and Hellenic Mathematical Society’s support, reflecting national interest in expanding quantum computing education at the high school level, with QPic now playing a significant role in this effort.
Notably, Greece is not the only country where QPic has supported national quantum education strategies. The inclusive nature of QPic has also been acknowledged by Elevate Quantum, the largest US quantum consortium prioritising quantum workforce development in the Mountain West. Under their mission to secure the region as a global epicenter for quantum, they have endorsed QPic as a key strategy for advancing equitable access to quantum science education.

As evidenced by the worldwide interest outlined above, the QPic experiment not only challenges assumptions about the prerequisites for Quantum Information Science and Technology, but, in doing so, highlights new pathways to quantum readiness. The findings open up new possibilities for early exposure to quantum, particularly at critical stages of education and career decision-making. As the next quantum revolution rapidly approaches, this raises important considerations for how to best prepare a skilled and adaptable workforce. 

There also is the more general goal of making STEM research more broadly attractive. One notable piece of feedback we received from  students was that “the course was so much fun”. With QPic-alike languages being developed for several other STEM areas, there is now a very promising prospect indeed to change the face of STEM subjects.

\section{Additional Information} 

Additional background, context, methodology and more details on the results reported in this paper can be found in the pre-experiment report~\cite{dundar2023quantum} and the extended and more technical version of this paper~\cite{dundarcoecke2025makingquantumworldaccessible}.  %Correspondence and requests for materials should be addressed to Caterina Puca (\href{mailto:caterina.puca@quantinuum.com}{caterina.puca@quantinuum.com}).

\section{Acknowledgements} 

We would like to extend our gratitude to the participating students, their supportive teachers, schools and families. Special thanks to the tutors who assisted tutorials during the experiment. We are also grateful to Karen Clayton for her help with logistical organisation. EMP was supported by European Office of Aerospace Research and Development (EOARD) grant FA8655-23-1-7220.

%\newpage
\bibliographystyle{plain}
\bibliography{mainNOW}

@misc{dundarcoecke2025makingquantumworldaccessible,
      title={Making the quantum world accessible to young learners through Quantum Picturalism: An experimental study}, 
      author={Selma Dündar-Coecke and Caterina Puca and Lia Yeh and Muhammad Hamza Waseem and Emmanuel M. Pothos and Thomas Cervoni and Sieglinde M. -L. Pfaendler and Vincent Wang-Maścianica and Peter Sigrist and Ferdi Tomassini and Vincent Anandraj and Ilyas Khan and Stefano Gogioso and Aleks Kissinger and Bob Coecke},
      year={2025, \emph{{arXiv preprint arXiv:2504.01013}}},
      journal={arXiv preprint arXiv:2504.01013}
}

@misc{kyma,
author = {},
title = {{KYMA} (Quantum Computing for Students)},
year = {2024},
note = {(accessed February 2025)}, 
howpublished = "\url{https://kyma.edu.gr/}"}

@article{orieux2016recent,
  title={Recent advances on integrated quantum communications},
  author={Orieux, Adeline and Diamanti, Eleni},
  journal={Journal of Optics},
  volume={18},
  number={8},
  pages={083002},
  year={2016},
  publisher={IOP Publishing}
}

@book{van2014quantum,
  title={Quantum networking},
  author={Van Meter, Rodney},
  year={2014},
  publisher={John Wiley \& Sons}
}

@article{dunjko2018machine,
  title={Machine learning \& artificial intelligence in the quantum domain: a review of recent progress},
  author={Dunjko, Vedran and Briegel, Hans J},
  journal={Reports on Progress in Physics},
  volume={81},
  number={7},
  pages={074001},
  year={2018},
  publisher={IOP Publishing}
}

@article{jeyaraman2024revolutionizing,
  title={Revolutionizing healthcare: the emerging role of quantum computing in enhancing medical technology and treatment},
  author={Jeyaraman, Naveen and Jeyaraman, Madhan and Yadav, Sankalp and Ramasubramanian, Swaminathan and Balaji, Sangeetha},
  journal={Cureus},
  volume={16},
  number={8},
  year={2024},
  publisher={Cureus}
}

@article{pal2024future,
  title={Future potential of quantum computing and simulations in biological science},
  author={Pal, Soumen and Bhattacharya, Manojit and Dash, Snehasish and Lee, Sang-Soo and Chakraborty, Chiranjib},
  journal={Molecular Biotechnology},
  volume={66},
  number={9},
  pages={2201--2218},
  year={2024},
  publisher={Springer}
}

@article{bauer2020quantum,
  title={Quantum algorithms for quantum chemistry and quantum materials science},
  author={Bauer, Bela and Bravyi, Sergey and Motta, Mario and Chan, Garnet Kin-Lic},
  journal={Chemical Reviews},
  volume={120},
  number={22},
  pages={12685--12717},
  year={2020},
  publisher={ACS Publications}
}

@article{cao2019quantum,
  title={Quantum chemistry in the age of quantum computing},
  author={Cao, Yudong and Romero, Jonathan and Olson, Jonathan P and Degroote, Matthias and Johnson, Peter D and Kieferov{\'a}, M{\'a}ria and Kivlichan, Ian D and Menke, Tim and Peropadre, Borja and Sawaya, Nicolas PD and others},
  journal={Chemical reviews},
  volume={119},
  number={19},
  pages={10856--10915},
  year={2019},
  publisher={ACS Publications}
}

@MISC{Alevel,
   author =       {UK Government},
   title =        {A level},
   month =        {8},
   year =         {2024},
   url =          {\url{https://educationhub.blog.gov.uk/category/a-level//}}
 }

@MISC{Geopolitics,
   author =       {The Geopolitics},
   title =        {The Geopolitical Rush Behind Quantum Technologies},
   month =        {9},
   year =         {2022},
   url =          {\url{https://thegeopolitics.com/the-geopolitical-rush-behind-quantum-technologies/}}
 }

@MISC{Forbes,
   author =       {Forbes},
   title =        {Quantum Computing Is Coming Faster Than You Think},
   month =        {11},
   year =         {2023},
   url =          {\url{https://www.forbes.com/sites/tiriasresearch/2023/11/28/quantum-computing-is-coming-faster-than-you-think/}}
 }

@misc{qdisccirctheory,
 author={Laakkonen, T. and Meichanetzidis, K. and  Coecke, B.},
 title={Quantum Algorithms for Compositional Text Processing},  
 note={arXiv preprint arXiv:2408.06061},
 year={2024},
 }

@misc{qdisccircexperiment,
 author={Duneau, T. and Bruhn, S. and Matos, G. and Laakkonen, T. and Santi, K. and Pearson, A. and Meichanetzidis, K. and  Coecke, B.},
 title={Scalable and interpretable quantum natural language processing: an implementation on trapped ions},  
 note={arXiv preprint arXiv:2409.08777},
 year={2024},
 }

@MISC{McKinsey,
   author =       {McKinsey \& Company},
   title =        {Steady progress in approaching the quantum advantage},
   month =        {4},
   year =         {2024},
   url =          {\url{https://www.mckinsey.com/capabilities/mckinsey-digital/our-insights/steady-progress-in-approaching-the-quantum-advantage}}
 }

@inproceedings{hadzihasanovic2018two,
  title={Two complete axiomatisations of pure-state qubit quantum computing},
  author={Hadzihasanovic, A. and Ng, K. F. and Wang, Q.},
  booktitle={Proceedings of the 33rd annual ACM/IEEE symposium on logic in computer science},
  pages={502--511},
  year={2018}
}

@inproceedings{dundar2023quantum,
  title={Quantum Picturalism: Learning Quantum Theory in High School},
  author={D{\"u}ndar-Coecke, S. and Yeh, L. and Puca, C. and Pfaendler, S. M-L and Waseem, M. H. and Cervoni, T. and Kissinger, A. and Gogioso, S. and Coecke, B.},
  booktitle={2023 IEEE International Conference on Quantum Computing and Engineering (QCE)},
  volume={3},
  pages={21--32},
  year={2023},
  organization={IEEE}
}

@book{von2005john,
  title={John von Neumann: selected letters},
  author={Von Neumann, J.},
  volume={27},
  year={2005},
  publisher={American Mathematical Society}
}

@misc{twinkle,
  title = {What is Angle?},
  author = {twinkle},
  Note = {https://www.twinkl.co.uk/teaching-wiki/angle}
}

@misc{khesinGraphicalQuantumCliffordencoder2023,
  title = {Graphical Quantum {{Clifford-encoder}} Compilers from the {{ZX}} Calculus},
  author = {Khesin, A. B. and Lu, J. Z. and Shor, P. W.},
  date = {2023-01-05},
  Note = {arXiv:2301.02356},
  eprint = {2301.02356},
  eprinttype = {arxiv},
  primaryclass = {quant-ph},
  publisher = {{arXiv}},
  doi = {10.48550/arXiv.2301.02356},
  url = {http://arxiv.org/abs/2301.02356}
}

@inproceedings{wille2022basis,
  title={The basis of design tools for quantum computing: arrays, decision diagrams, tensor networks, and ZX-calculus},
  author={Wille, Robert and Burgholzer, Lukas and Hillmich, Stefan and Grurl, Thomas and Ploier, Alexander and Peham, Tom},
  booktitle={Proceedings of the 59th ACM/IEEE Design Automation Conference},
  pages={1367--1370},
  year={2022}
}

@article{litinski2022active,
  title={Active volume: An architecture for efficient fault-tolerant quantum computers with limited non-local connections},
  author={Litinski, D. and Nickerson, N.},
  journal={arXiv preprint arXiv:2211.15465},
  year={2022}
}

@book{QiP,
  title={Quantum in Pictures},
  author={Coecke, B. and Gogioso, S.},
  year={2022},
  publisher={Quantinuum}
}

@article{kartsaklis2021lambeq,
      title={lambeq: An Efficient High-Level {P}ython Library for Quantum {NLP}}, 
      author={D. Kartsaklis and I. Fan and R. Yeung and A. Pearson and R. Lorenz and A. Toumi and G. de Felice and K. Meichanetzidis and S. Clark and B. Coecke},
      year={2021},
      journal={arXiv preprint arXiv:2110.04236}
}

@article{Gidney2019,
  doi = {10.22331/q-2019-04-30-135},
  url = {https://doi.org/10.22331/q-2019-04-30-135},
  title = {Efficient magic state factories with a catalyzed {$|CCZ\rangle$} to {$2|T\rangle$} transformation},
  author = {Gidney, C. and Fowler, A. G.},
  journal = {{Quantum}},
  issn = {2521-327X},
  publisher = {{Verein zur F{\"{o}}rderung des Open Access Publizierens in den Quantenwissenschaften}},
  volume = {3},
  pages = {135},
  month = apr,
  year = {2019}
}

@article{KissingerTcount,
  title = {Reducing the number of non-{C}lifford gates in quantum circuits},
  author = {Kissinger, A. and van de Wetering, J.},
  journal = {Phys. Rev. A},
  volume = {102},
  issue = {2},
  pages = {022406},
  numpages = {10},
  year = {2020},
  month = {8},
  publisher = {American Physical Society},
  doi = {10.1103/PhysRevA.102.022406},
  url = {https://link.aps.org/doi/10.1103/PhysRevA.102.022406}
}

@article{clifford-simp,
  title={Graph-theoretic Simplification of Quantum Circuits with the {ZX}-calculus},
  author={Duncan, R. and Kissinger, A. and Perdrix, S. and Van De Wetering, J.},
  journal={Quantum},
  volume={4},
  pages={279},
  year={2020},
  publisher={Verein zur F{\"o}rderung des Open Access Publizierens in den Quantenwissenschaften}
}

@article{vilmart2018near,
  title={A near-optimal axiomatisation of {ZX}-calculus for pure qubit quantum mechanics},
  author={Vilmart, R.},
  journal={arXiv preprint arXiv:1812.09114},
  year={2018}
}

@inproceedings{de2020fast,
  title={Fast and Effective Techniques for T-Count Reduction via Spider Nest Identities},
  author={de Beaudrap, N. and Bian, X. and Wang, Q.},
  booktitle={15th Conference on the Theory of Quantum Computation, Communication and Cryptography (TQC 2020)},
  year={2020},
  organization={Schloss Dagstuhl-Leibniz-Zentrum f{\"u}r Informatik}
}

@article{Nature,
  title={Grammar-Aware Question-Answering on Quantum Computers},
  author={Meichanetzidis, K. and Toumi, A. and de Felice, G. and Coecke, B.},
  journal={arXiv preprint arXiv:2012.03756},
  year={2020}
}

@article{cowtan2019phase,
  title={Phase Gadget Synthesis for Shallow Circuits},
  author={Cowtan, A. and Dilkes, S. and Duncan, R. and Simmons, W. and Sivarajah, S.},
  journal={arXiv preprint arXiv:1906.01734},
  year={2019}
}

@article{gogioso2017fully,
  title={Fully graphical treatment of the quantum algorithm for the Hidden Subgroup Problem},
  author={Gogioso, S. and Kissinger, A.},
  journal={arXiv preprint arXiv:1701.08669},
  year={2017}
}

@book{CKbook,
	Author = {B. Coecke and A. Kissinger},
	Publisher = {Cambridge University Press},
	Title = {Picturing Quantum Processes. A First Course in Quantum Theory and Diagrammatic Reasoning},
	doi= {10.1017/9781316219317},
	Year = {2017}}

@inproceedings{AC1,
	Author = {S. Abramsky and B. Coecke},
	Booktitle = {Proceedings of the 19th Annual IEEE Symposium on Logic in Computer Science (LICS)},
	Note = {{a}rXiv:quant-ph/0402130},
	Pages = {415-425},
	Title = {A categorical semantics of quantum protocols},
	Year = {2004}}

@article{Benabou,
	Author = {Benabou, J.},
	Journal = {Comptes Rendus des S\'eances de l'Acad\'emie des Sciences. Paris},
	Pages = {1887--1890},
	Title = {Categories avec multiplication},
	Volume = {256},
	Year = {1963}}

@article{CarboniWalters,
	Author = {Carboni, A. and Walters, R. F. C.},
	Journal = {Journal of Pure and Applied Algebra},
	Pages = {11-32},
	Title = {Cartesian bicategories {I}},
	Volume = {49},
	Year = {1987}}

@inproceedings{Kindergarten,
	Author = {B. Coecke},
	Booktitle = {Quantum Theory: Reconsiderations of the Foundations III},
	Editor = {A. Khrennikov},
	Note = {{a}rXiv:quant-ph/0510032},
	Pages = {81--98},
	Publisher = {AIP Press},
	Title = {Kindergarten quantum mechanics},
	Year = 2005}

@article{ContPhys,
	Author = {B. Coecke},
	Journal = {Contemporary Physics},
	Note = {{a}rXiv:0908.1787},
	Pages = {59-83},
	Title = {Quantum picturalism},
	Volume = {51},
	Year = {2009}}

@inproceedings{CD1,
	Author = {B. Coecke and R. Duncan},
	Booktitle = {Proceedings of the 37th International Colloquium on Automata, Languages and Programming (ICALP)},
	Series = {Lecture Notes in Computer Science},
	Title = {Interacting quantum observables},
	Year = {2008}}

@article{CD2,
	Author = {B. Coecke and R. Duncan},
	Journal = {New Journal of Physics},
	Note = {{arXiv:quant-ph/09064725}},
	Pages = {043016},
	Title = {Interacting quantum observables: categorical algebra and diagrammatics},
	Volume = {13},
	Year = {2011}}

@inproceedings{CDKZ,
	Author = {Coecke, B. and Duncan, R. and Kissinger, A. and Wang, Q.},
	Booktitle = {Proceedings of the 27th Annual IEEE Symposium on Logic in Computer Science (LICS)},
	Note = {arXiv:1203.4988},
	Title = {Strong complementarity and non-locality in categorical quantum mechanics},
	Year = {2012}}

@incollection{CPaqPav,
	Author = {B. Coecke and {\'E}. O. Paquette and D. Pavlovi{\'c}},
	Booktitle = {Semantic Techniques in Quantum Computation},
	Editor = {S. Gay and I. Mackie},
	Note = {{a}rXiv:0904.1997},
	Pages = {29--69},
	Publisher = {Cambridge University Press},
	Title = {{Classical and quantum structuralism}},
	Year = {2010}}

@inproceedings{CPer,
	Author = {B. Coecke and S. Perdrix},
	Booktitle = {Proceedings of the 19th EACSL Annual Conference on Computer Science Logic (CSL)},
	Note = {Extended version: {a}rXiv:1004.1598},
	Pages = {230-244},
	Series = {Lecture Notes in Computer Science},
	Title = {Environment and classical channels in categorical quantum mechanics},
	Volume = {6247},
	Year = {2010}}

@article{KellyLaplaza,
	Author = {G.~M. Kelly and M. L. Laplaza},
	Journal = {Journal of Pure and Applied Algebra},
	Pages = {193-213},
	Title = {Coherence for compact closed categories},
	Volume = {19},
	Year = {1980}}

@article{Lack,
	Author = {S. Lack},
	Journal = {Theory and Applications of Categories},
	Pages = {147-163},
	Title = {Composing {PROP}s},
	Volume = {13},
	Year = {2004}}

@incollection{Penrose,
	Author = {R.~Penrose},
	Booktitle = {Combinatorial Mathematics and its Applications},
	Pages = {221--244},
	Publisher = {Academic Press},
	Title = {Applications of negative dimensional tensors},
	Year = 1971}

@article{Redei1,
	Author = {Redei, M.},
	Journal = {Studies in History and Philosophy of Modern Physics},
	Number = {4},
	Pages = {493--510},
	Title = {Why {J}ohn von {N}eumann did not like the {H}ilbert Space formalism of quantum mechanics (and what he liked instead)},
	Volume = {27},
	Year = {1996}}

@article{SelingerCPM,
	Author = {P. Selinger},
	Journal = {Electronic Notes in Theoretical Computer Science},
	Pages = {139--163},
	Publisher = {Elsevier},
	Title = {Dagger compact closed categories and completely positive maps},
	Volume = {170},
	Year = {2007}}

@book{vN,
	Author = {von Neumann, J.},
	Note = {Translation, {\it Mathematical foundations of quantum mechanics}, Princeton University Press, 1955.},
	Publisher = {Springer-Verlag},
	Title = {Mathematische grundlagen der quantenmechanik},
	Year = {1932}}

@article{vandaele2024qubitcount,
    author = {Vandaele, Vivien},
    title = {{Qubit-count optimization using ZX-calculus}},
    year = {2024},
    journal = {arXiv preprint arXiv:2407.10171}
}

@inproceedings{borgna2021hybrid,
    author = {Borgna, Agust{\'i}n and Perdrix, Simon and Valiron, Beno{\^i}t},
    title = {{Hybrid quantum-classical circuit simplification with the ZX-calculus}},
    year = {2021},
    booktitle = {Programming Languages and Systems},
    editor = {Oh, Hakjoo},
    pages = {121--139},
    publisher = {Springer International Publishing},
    address = {Cham},
    doi = {10.1007/978-3-030-89051-3_8},
    video = {https://www.youtube.com/watch?v=-Frd8w-hTD0}
}

@inproceedings{gogioso2023annealing,
    author = {Gogioso, Stefano and Yeung, Richie},
    title = {Annealing Optimisation of Mixed ZX Phase Circuits},
    year = {2023},
    booktitle = {Proceedings 19th International Conference on Quantum Physics and Logic, Wolfson College, Oxford, UK, 27 June - 1 July 2022},
    editor = {Gogioso, Stefano and Hoban, Matty},
    series = {Electronic Proceedings in Theoretical Computer Science},
    volume = {394},
    pages = {415-431},
    publisher = {Open Publishing Association},
    doi = {10.4204/EPTCS.394.20},
    video = {https://youtu.be/WZgDVPXo7WQ?t=2706}
}

@article{holker2023causal,
    author = {Holker, Calum},
    title = {{Causal flow preserving optimisation of quantum circuits in the ZX-calculus}},
    year = {2023},
    journal = {arXiv preprint arXiv:2312.02793},
    video = {https://www.youtube.com/watch?v=9wO_-QcSAaM}
}

@inproceedings{kissinger2022classical,
    author = {Kissinger, Aleks and van de Wetering, John and Vilmart, Renaud},
    title = {{Classical Simulation of Quantum Circuits with Partial and Graphical Stabiliser Decompositions}},
    year = {2022},
    booktitle = {17th Conference on the Theory of Quantum Computation, Communication and Cryptography (TQC 2022)},
    editor = {Le Gall, Fran\c{c}ois and Morimae, Tomoyuki},
    series = {Leibniz International Proceedings in Informatics (LIPIcs)},
    volume = {232},
    pages = {5:1--5:13},
    publisher = {Schloss Dagstuhl -- Leibniz-Zentrum f{\"u}r Informatik},
    address = {Dagstuhl, Germany},
    isbn = {978-3-95977-237-2},
    issn = {1868-8969},
    doi = {10.4230/LIPIcs.TQC.2022.5},
    video = {https://youtu.be/y3LIDcgrc1k?t=1394}
}

@article{codsi2022classically,
    author = {Codsi, Julien and van de Wetering, John},
    title = {{Classically Simulating Quantum Supremacy IQP Circuits trough a Random Graph Approach}},
    year = {2022},
    journal = {arXiv preprint arXiv:2212.08609}
}

@article{ufrecht2023cutting,
    author = {Ufrecht, Christian and Periyasamy, Maniraman and Rietsch, Sebastian and Scherer, Daniel D. and Plinge, Axel and Mutschler, Christopher},
    title = {Cutting multi-control quantum gates with {ZX} calculus},
    year = {2023},
    journal = {{Quantum}},
    volume = {7},
    month = {10},
    pages = {1147},
    publisher = {{Verein zur F{\"{o}}rderung des Open Access Publizierens in den Quantenwissenschaften}},
    issn = {2521-327X},
    doi = {10.22331/q-2023-10-23-1147},
    url = {https://doi.org/10.22331/q-2023-10-23-1147},
    video = {https://www.youtube.com/watch?v=YPtEIuapWww}
}

@article{cam2023speeding,
    author = {Cam, Tristan and Martiel, Simon},
    title = {{Speeding up quantum circuits simulation using ZX-Calculus}},
    year = {2023},
    journal = {arXiv preprint arXiv:2305.02669}
}

@inproceedings{KoziellPipe2024Simulation,
    author = {Koziell-Pipe, Alex and Yeung, Richie and Sutcliffe, Matthew},
    title = {{Towards Faster Quantum Circuit Simulation Using Graph Decompositions, GNNs and Reinforcement Learning}},
    year = {2024},
    booktitle = {38th Conference on Neural Information Processing Systems (NeurIPS 2024)}
}

@article{kissinger2022grok1,
    author = {Kissinger, Aleks},
    title = {{Phase-free ZX diagrams are CSS codes (...or how to graphically grok the surface code)}},
    year = {2022},
    journal = {arXiv preprint arXiv:2204.14038},
    video = {https://www.youtube.com/watch?v=VVhBGpCMlwA}
}

@article{bombin2023unifying,
    author = {Bombin, Hector and Litinski, Daniel and Nickerson, Naomi and Pastawski, Fernando and Roberts, Sam},
    title = {{Unifying flavors of fault tolerance with the ZX calculus}},
    year = {2023},
    journal = {arXiv preprint arXiv:2303.08829}
}

@article{wu2023zxcalculus,
    author = {Wu, Zipeng and Cheng, Song and Zeng, Bei},
    title = {{A ZX-Calculus Approach to Concatenated Graph Codes}},
    year = {2023},
    journal = {arXiv preprint arXiv:2304.08363},
    video = {https://www.youtube.com/watch?v=mbqx0OudY5c}
}

@inproceedings{townsend-teague2023floquetifying,
    author = {Townsend-Teague, Alex and Magdalena de la Fuente, Julio and Kesselring, Markus},
    title = {{Floquetifying the Colour Code}},
    year = {2023},
    booktitle = {Proceedings of the Twentieth International Conference on Quantum Physics and Logic, Paris, France, 17-21st July 2023},
    editor = {Mansfield, Shane and Val\^iron, Benoit and Zamdzhiev, Vladimir},
    series = {Electronic Proceedings in Theoretical Computer Science},
    volume = {384},
    pages = {265-303},
    publisher = {Open Publishing Association},
    doi = {10.4204/EPTCS.384.14},
    video = {https://www.youtube.com/watch?v=pP1DaKxlC_8}
}

@article{Huang2023,
   title={Graphical CSS Code Transformation Using ZX Calculus},
   volume={384},
   ISSN={2075-2180},
   url={http://dx.doi.org/10.4204/EPTCS.384.1},
   DOI={10.4204/eptcs.384.1},
   journal={Electronic Proceedings in Theoretical Computer Science},
   publisher={Open Publishing Association},
   author={Huang, Jiaxin and Li, Sarah Meng and Yeh, Lia and Kissinger, Aleks and Mosca, Michele and Vasmer, Michael},
   year={2023},
   month=aug, pages={1–19}
}

@inproceedings{kissinger2024grok2,
    author = {Kissinger, Aleks and van de Wetering, John},
    title = {{Scalable Spider Nests (...Or How to Graphically Grok Transversal Non-Clifford Gates)}},
    year = {2024},
    booktitle = {Proceedings of the 21st International Conference on Quantum Physics and Logic, Buenos Aires, Argentina, July 15-19, 2024},
    editor = {D\'iaz-Caro, Alejandro and Zamdzhiev, Vladimir},
    series = {Electronic Proceedings in Theoretical Computer Science},
    volume = {406},
    pages = {79-95},
    publisher = {Open Publishing Association},
    doi = {10.4204/EPTCS.406.4}
}

@article{burton2024genons,
    author = {Burton, Simon and Durso-Sabina, Elijah and C. Brown, Natalie},
    title = {{Genons, Double Covers and Fault-tolerant Clifford Gates}},
    year = {2024},
    journal = {arXiv preprint arXiv:2406.09951}
}

@article{boriskhesin2024equivalence,
    author = {Boris Khesin, Andrey and Li, Alexander},
    title = {{Equivalence Classes of Quantum Error-Correcting Codes}},
    year = {2024},
    journal = {arXiv preprint arXiv:2406.12083}
}

@article{magdalenadelafuente2024xyz,
    author = {C. Magdalena de la Fuente, Julio and Old, Josias and Townsend-Teague, Alex and Rispler, Manuel and Eisert, Jens and M{\"u}ller, Markus},
    title = {{The XYZ ruby code: Making a case for a three-colored graphical calculus for quantum error correction in spacetime}},
    year = {2024},
    journal = {arXiv preprint arXiv:2407.08566}
}

@article{rodatz2024floquetifying,
    author = {Rodatz, Benjamin and Po{\'o}r, Boldizs{\'a}r and Kissinger, Aleks},
    title = {{Floquetifying stabiliser codes with distance-preserving rewrites}},
    year = {2024},
    journal = {arXiv preprint arXiv:2410.17240}
}

%\printbibliography

\appendix

\end{document}